\documentclass[multphys,vecphys,norunningheads]{svmult}

\usepackage{makeidx}         % allows index generation
\usepackage{graphicx}        % standard LaTeX graphics tool
\usepackage{multicol}        % used for the two-column index
\usepackage[bottom]{footmisc}% places footnotes at page bottom
\makeindex             % used for the subject index
\newcommand{\magh}{$\gamma$-Fe$_2$O$_3$}
\newcommand{\beq}{\begin{equation}}
\newcommand{\eeq}{\end{equation}}
\newcommand{\bea}{\begin{eqnarray}}
\newcommand{\eea}{\end{eqnarray}}

\usepackage{amssymb}
\usepackage{graphicx}

\begin{document}
\DeclareGraphicsRule{.eps.gz}{eps}{.eps.bb}{` gunzip -c #1}
\title*{FROM FINITE-SIZE AND SURFACE EFFECTS TO GLASSY BEHAVIOUR IN FERRIMAGNETIC NANOPARTICLES}
\titlerunning{From finite-size and surface effects to glassy behaviour}
%Use \titlerunning{Short Title} for an abbreviated version of
\author{Am\'{\i}lcar Labarta, Xavier Batlle and \`{O}scar Iglesias}
\institute{\it {Departament de F\'{\i}sica Fonamental
Universitat de Barcelona, Diagonal 647, 08028 Barcelona, Spain}\\
\texttt{amilcar@ffn.ub.es, xavier@ffn.ub.es, oscar@ffn.ub.es}
}
\maketitle

This chapter is aimed at studying the anomalous magnetic properties (glassy behaviour) observed at low temperatures in nanoparticles of ferrimagnetic oxides. 
This topic is discussed both from numerical results and experimental data. 
Ferrimagnetic fine particles show most of the features of glassy systems due to the random distribution of anisotropy axis, interparticle interactions and surface effects. Experiments have shown that the hysteresis loops display high closure fields with high values of the differential susceptibility. Low magnetisation as compared to bulk, shifted loops after field-cooling, high-field irreversibilities between zero-field and field cooling processes and ageing phenomena in the time-dependence of the magnetisation, are also observed. 
This phenomenology indicates the existence of some kind of freezing phenomenon arising from a complex hierarchy of the energy levels, whose origin is currently under discussion. 
Two models have been proposed to account for it: i) the existence of a spin-glass state at the surface of the particle which is coupled to the particle core through an exchange field; and ii) the collective behaviour induced by interparticle interactions. In real systems, both contributions simultaneously occur, being difficult to distinguish their effects. In contrast, numerical simulations allow us to build a model just containing the essential ingredients to study solely one of two phenomena.

%\begin{keywords}
%Ferrimagnets,...
%05.10 Ln, 75.40 Cx, 75.40.Mg, 75.50 Gg, 75.50 Tf, 75.60 Ej
%\end{keywords}

\section{Frustration in ferrimagnetic oxides}
\label{Frustration_Sec}

Ferrimagnetic oxides are one kind of magnetic systems in which there exist at least two inequivalent sublattices for the magnetic ions. The antiparallel alignment between these sublattices (ferrimagnetic ordering) may occur provided the intersublattice exchange interactions are antiferromagnetic (AF) and some requirements concerning the signs and strengths of the intrasublattice interactions are fulfilled. 
Since usually in ferrimagnetic oxides the magnetic cations are surrounded by bigger oxygen anions (almost excluding the direct overlap between cation orbitals) magnetic interactions occur via indirect superexchange mediated by the $p$ oxygen orbitals. It is well-known that the sign of these superexchange interactions depends both on the electronic structure of the cations and their geometrical arrangement \cite{Krupickabook82}. In most ferrimagnetic oxides, the crystallographic and electronic structure give rise to antiferromagnetic inter- and intrasublattice competing interactions.
Consequently, in these cases not all magnetic interactions can be fulfilled, and in spite of the collinear alignment of the spins, some degree of magnetic frustration exists. 

\begin{figure}[tp] 
\centering 
\includegraphics[width= 0.8\textwidth]{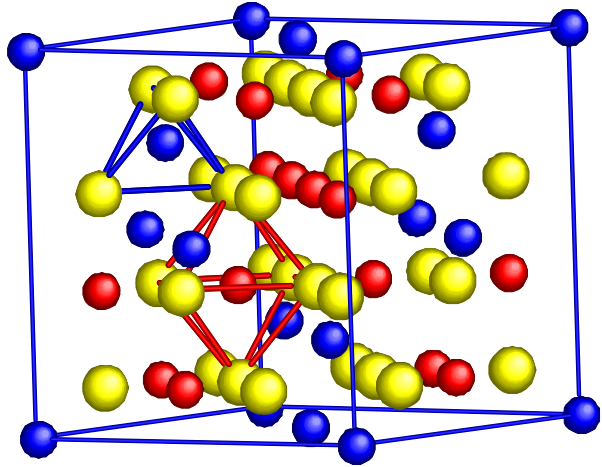}
\caption{Unit cell of maghemite. The magnetic Fe$^{3+}$ ions occupying the two 
sublattices, in different coordination with the O$^{2-}$ ions (light-grey colour), are 
coloured in black (T sublattice, tetrahedric coordination) and in dark grey  
(O sublattice, octahedric coordination). 
}  
\label{Fe2O3_bw_fig} 
\end{figure}

The most common crystallographic structures for ferrimagnetic oxides are hexagonal ferrites, garnets, and spinels, all of them having intrinsic geometrical frustration when interactions are all antiferromagnetic \cite{Andersonpr56}. 
A typical ferrimagnetic oxide with spinel structure, having a variety of technological applications, is $\gamma$-Fe$_2$O$_3$ (maghemite), in which the magnetic Fe$^{3+}$ ions with spin $S= 5/2$ are disposed in two sublattices with different coordination with the O$^{2-}$ ions. Each unit cell (see Fig. \ref{Fe2O3_bw_fig}) has 8 tetrahedric (T), 16 octahedric (O) sites, and one sixth  
of the O sites have randomly distributed vacancies to achieve neutrality charge.  
The T sublattice has larger coordination than O, thus, while the spins in the T  
sublattice have $N_{TT}=4$ nearest neighbours in T sites and $N_{TO}=12$ in O sites,  
the spins in the O sublattice have $N_{OO}=6$ nearest neighbours in O and  
$N_{OT}=6$ in T. The values of the nearest-neighbour exchange constants are
\cite{Kodamaprb99,Kachkachiepj00,Iglesiasprb01}: $J_{TT}=-21$ K, $J_{OO}= -8.6$ K, $J_{TO}= -28.1$ K. So, the local magnetic energy balance favours ferrimagnetic order, with spins in each sublattice ferromagnetically aligned and antiparallel intrasublattice alignment.

The substitution of magnetic ions by vacancies or the presence of broken bonds between both sublattices destabilises the collinear arrangement of the spins inducing spin canting in the sublattice with less degree of substitution, a fact that has been observed in many ferrimagnetic oxides with substitution by non-magnetic cations  \cite{Gellerjap66,Rosencwaigcjp70,Pattonjpc83,Albaneseap76,Obradorsjpc86,Batllejap93}. 
It is worth noting that when the ferrimagnetic oxide is in the form of small particles, the structural modifications at the boundaries of the particle (vacancies, broken bonds and modified exchange interactions) may induce enough frustration so as to destabilise the ferrimagnetic order at the surface layer giving rise to different canted magnetic structures.

In this chapter, we will discuss how these canted spins at the surface of the particle may freeze giving rise to a glassy state below a certain temperature.  
%----------------------------FIG.0--------------------------------------------- 
%----------------------------FIG.0--------------------------------------------- 
%%%%%%%%%%%%%%%%%%%%%%%%%%%%%%%%%%%%%%%%%%%%%%%%%%%%%%%%%%%%%%%%%%%%%%%%%%%%%%%%%%%%%%%%%%
\section{Glassy behaviour in ferrimagnetic nanoparticles}
The assemblies of fine magnetic particles with large packing fractions and/or nanometric sizes show most of the features which are characteristic of glassy systems (for a recent review see Ref.  \cite{Batllejpd02}). This glassy behaviour results from a complex interplay between surface and finite-size effects, interparticle interactions and the random distribution of anisotropy axis throughout the system. 
In many cases, these contributions are mixed and in competition, and their effects are thus generally blurred. However, several works propose that glassy behaviour in fine particle systems is mostly attributed to the frustration induced by strong magnetic interactions among particles with a random distribution of anisotropy axes \cite{Mamiyaprl98,Morupepl94,Morupprb95}, although some experimental and numerical results suggest the dominant role of the spin disorder at the surface layer in ferrimagnetic oxides 
\cite{Parkerprb93,Kodamaprl96,Coeyprl71,Pankhurstprl91,Respaudprb98,Martinezprl98}.

\subsection{Anomalous magnetic properties at low temperature}
\label{Anomalous_Sec}
The most important features characterising the glassy state in fine-particle systems include the flattening of the field-cooling (FC) susceptibility \cite{Preneie93}, the existence of high-field irreversibility in the magnetisation curves and between the field and zero-field-cooling (ZFC) susceptibilities \cite{Kodamaprb99,Martinezprl98,Montseprb99,Kodamaprl97},
the occurrence of shifted loops after FC \cite{Kodamaprl97,Moralescm99,Montseprb99,Troncjm03},
the increase in the magnetic viscosity \cite{Dormannjm98}, the critical slowing down observed by ac susceptibility \cite{Djurbergprl97}, the occurrence of ageing phenomena \cite{Dormannjm98,Jonssonprl95,Djurbergprl97,Mamiyaprl99,Jonssonprb98,Jonssonprb00} and the rapid increase of the non-linear susceptibility as the blocking temperature is approached from above \cite{Dormannjm98}. 
Even though these features are not associated with the occurrence of a true spin-glass transition in interacting fine-particle systems, some authors claim that they may be indicative of the existence of some kind of collective state \cite{Mamiyaprl98,Morupepl94,Morupprb95,Dormannjm98}. Here we discuss an example of a complete magnetic study in a nanocrystalline barium ferrite which is representative of this glassy behaviour.

\subsubsection{\it Sample characterisation}
The phenomenology of the glassy state in strong interacting fine particles is illustrated through the study of the magnetic properties of nanocrystalline BaFe$_{10.4}$Co$_{0.8}$Ti$_{0.8}$O$_{19}$ \cite{Batllejap93}. 
\textit{M}-type barium ferrites have been studied for a long time because of their technological applications \cite{Sharrockie86,Fujiwaraie87,Sharrockie89,Kryderjm90}, such as microwave devices, permanent magnets, and high-density magnetic and magneto-optic recording media, as well as their large pure research interest \cite{Kojima82,Labartaprb92}. The compounds obtained by cationic substitution of the pure BaFe$_{12}$O$_{19}$ ferrite display a large variety of magnetic properties and structures, which go from collinear ferrimagnetism to spin-glass-like behaviour \cite{Kojima82,Batllejap91,Labartaprb92}, depending on the degree of frustration introduced by cationic substitution. In particular, the BaFe$_{10.4}$Co$_{0.8}$Ti$_{0.8}$O$_{19}$ compound seems to be ideal for perpendicular magnetic recording \cite{Gornertem91}, since the Co$^{2+}$-Ti$^{4+}$ doping scheme reduces sharply the high values of the coercive field of the pure compound, which precludes their technological applications. For this composition the magnetic structure is still ferrimagnetic, the Curie temperature is well above room temperature and the coercive field is reduced to usual values for magnetic recording technologies. 

Nanocrystalline BaFe$_{10.4}$Co$_{0.8}$Ti$_{0.8}$O$_{19}$ samples were prepared by the glass crystallisation method (GCM). The GCM is characterised by the presence of homogenised melt fluxes of Fe$_2$O$_3$, BaO, B$_2$O$_3$ and the corresponding oxides of the doping cations (Ti$_2$O and CoO for the Co-Ti substitution) at about 1300 $^\circ$C, which are amorphised by rapid quenching in a two-roller equipment. Annealing the glass flakes above 550 $^\circ$C leads to the nucleation and growth of the borate and hexaferrite phases. Barium ferrite particles crystallise to suitable sizes during this treatment and may be isolated by dissolving the matrix in dilute acetic acid in an ultrasonic field. After centrifugation, washing and drying, a fine powder of \textit{M}-type doped barium ferrite particles with the desired stoichiometry is obtained. A more detailed description of this method can be found in Refs. \cite{Gornertem91,Gornertjm91}.
%----------------------------FIG.8--------------------------------------------- 
\begin{figure}[tp] 
\centering 
\includegraphics[width= 1.0\textwidth]{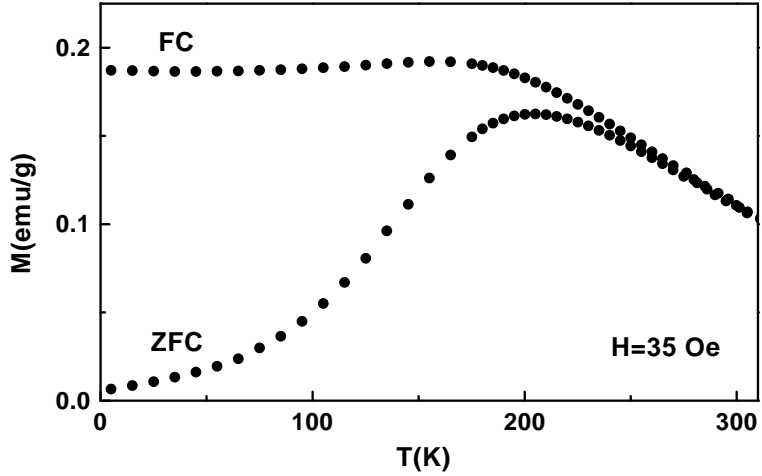}
\caption{Zero-field-cooling and field-cooling magnetisation
as a function of temperature measured at a field of 35 Oe for a nanocrystalline 
BaFe$_{10.4}$Co$_{0.8}$Ti$_{0.8}$O$_{19}$ sample.
}  
\label{Fig1a_fig} 
\end{figure}
%----------------------------FIG.8--------------------------------------------- 

In order to maximise interparticle interactions the sample was studied in powder form. X-ray diffraction data showed very broad peaks and the fitting of the whole spectra to the \textit{M}-type structure demonstrated the platelet-like morphology of the particles. The particle size distribution as determined from transmission electron microscopy (TEM) and X-ray diffraction was log normal with a mean platelet diameter of 10.2 nm, $\sigma$=0.48 and an aspect ratio of 4 (mean volume of $10^5$ nm$^3$). TEM also confirmed the platelet-like shape and showed a certain tendency of the particles to pile up producing stacks along the perpendicular direction to the (001) face of the platelet, which corresponds to the easy axis. TEM data showed the existence of large agglomerates composed of stacks, particle aggregates and quasispherical conglomerates, as expected from the value of the aspect ratio of these particles. The powder was mixed with a glue in order to avoid particle rotation towards the applied field axis during magnetic measurements.
%----------------------------FIG.9--------------------------------------------- 
\begin{figure}[tp] 
\centering 
\includegraphics[width= 0.9\textwidth]{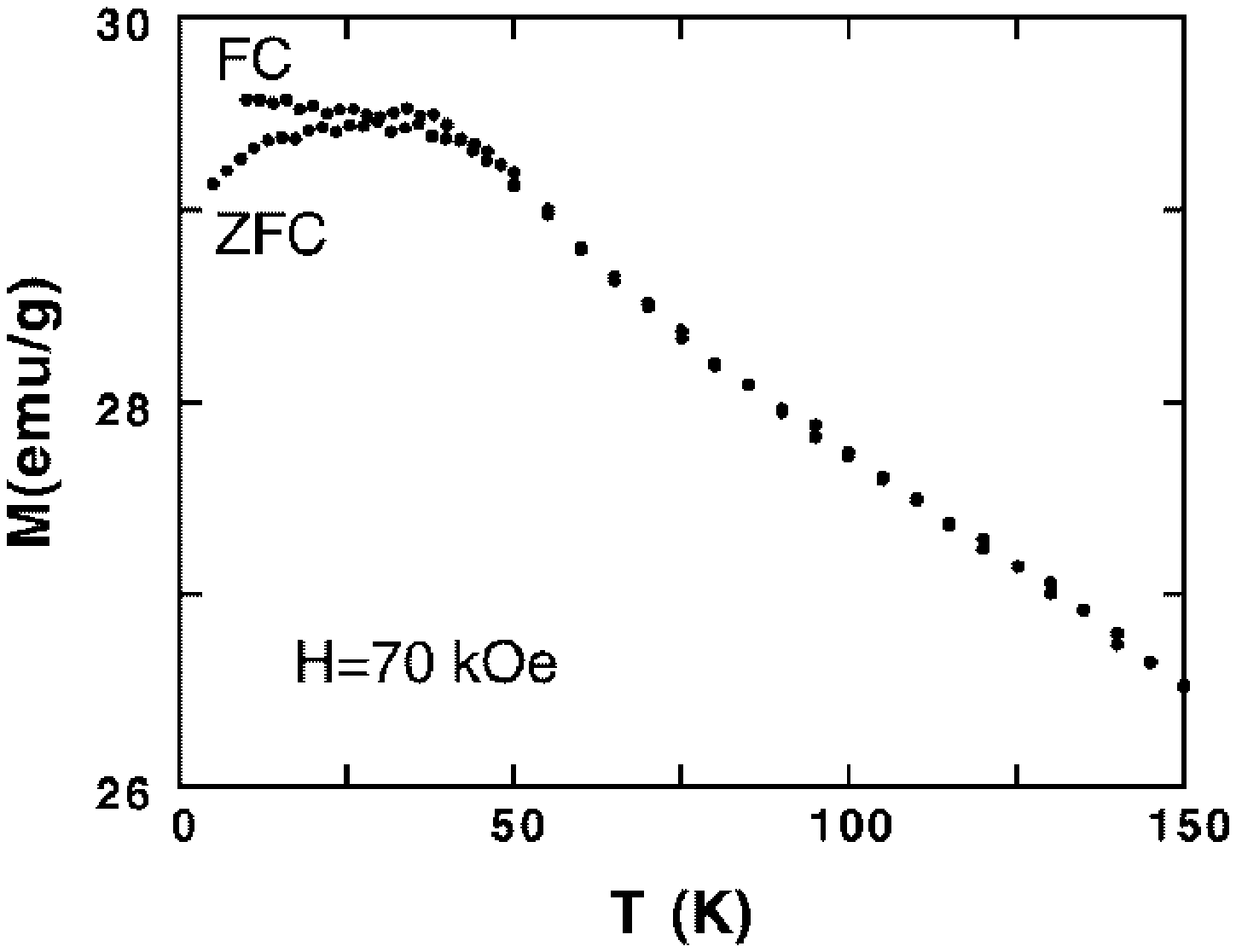}
\caption{Zero-field-cooling and field-cooling magnetisation
as a function of temperature measured at a field of 70 kOe for a nanocrystalline 
BaFe$_{10.4}$Co$_{0.8}$Ti$_{0.8}$O$_{19}$ sample.
}  
\label{Fig2a_fig} 
\end{figure}
%----------------------------FIG.9--------------------------------------------- 

\subsubsection{\it Magnetic properties}
Magnetisation measurements were carried out with a SQUID magnetometer under magnetic fields up to 55 kOe. High-field magnetisation measurements (up to 240 kOe) were performed at the Grenoble High Field Laboratory, using a water-cooled Bitter magnet with an extraction magnetometer. The temperature dependence of the AC magnetic susceptibility was recorded at frequencies between 10 Hz and 1 kHz, applying an AC magnetic field of 1 Oe after cooling the sample at zero field from room temperature.

The zero-field-cooling and field-cooling curves measured at 35 Oe displayed all the typical features of an assembly of small magnetic particles with a distribution of energy barriers (see Fig. \ref{Fig1a_fig}). The ZFC curve showed a broad maximum located at ca. $T_M= 205\pm 5$ K, which originated from both blocking and freezing processes, the latter due to magnetic frustration induced by interparticle interactions. 
%
%----------------------------FIG.10--------------------------------------------- 
\begin{figure}[tp] 
\centering 
\includegraphics[width= 1.0\textwidth]{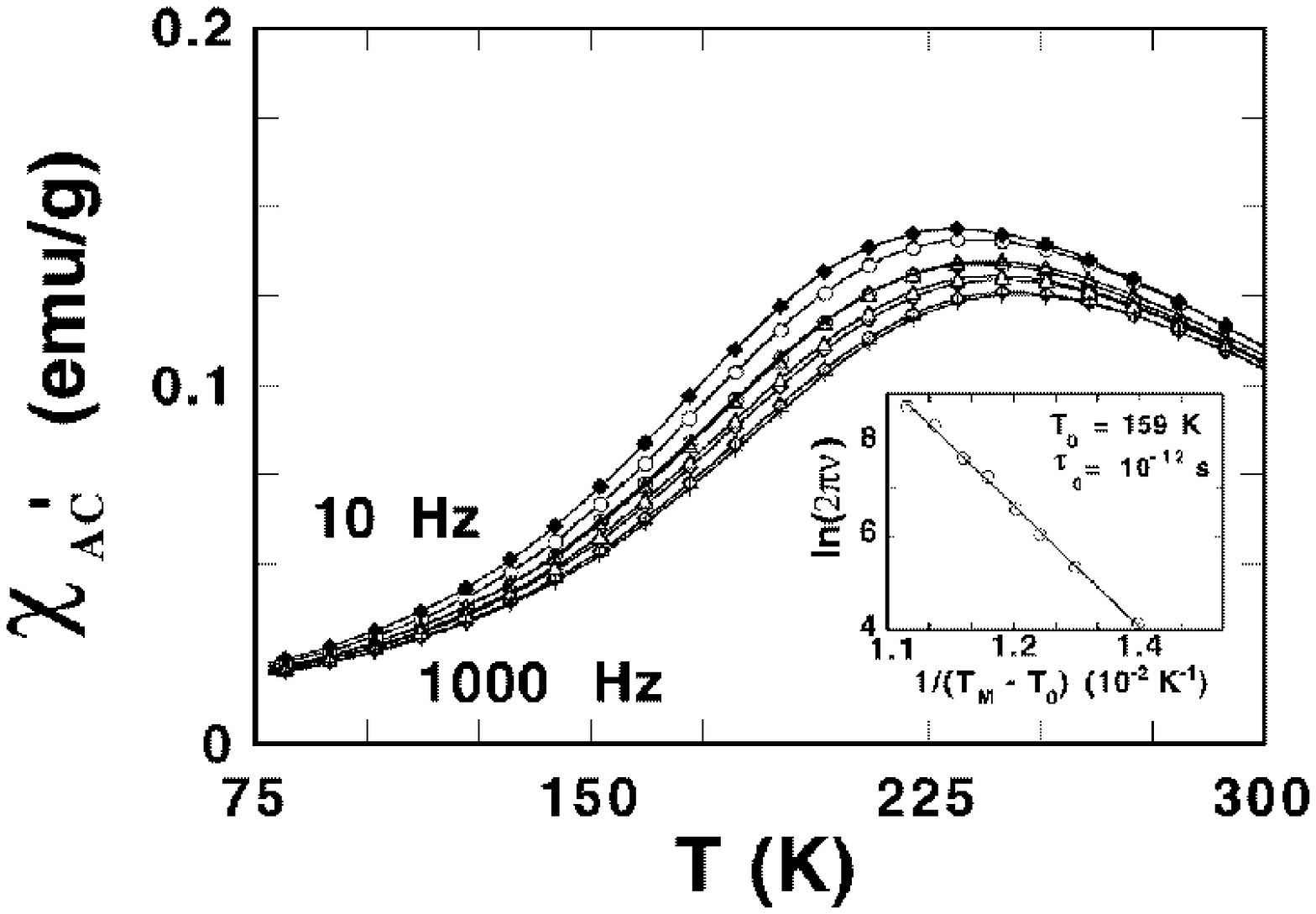}
\caption{Temperature dependence of the in-phase component
of the AC susceptibility ($\chi_{AC}'$) of a nanocrystalline 
BaFe$_{10.4}$Co$_{0.8}$Ti$_{0.8}$O$_{19}$ sample measured at different
frequencies in the range between 10 Hz and 1 kHz. Inset:
logarithm of the measuring frequency as a function of the
reciprocal of the difference between the temperature of the
peak and $T_0$.
}  
\label{Fig3a_fig} 
\end{figure}
%----------------------------FIG.10---------------------------------------------
The FC was very flat below $T_M$ , in comparison with the monotonously increasing behaviour characteristic of noninteracting systems, which reinforced the existence of strong interactions among particles. 
Moreover, in the superparamagnetic (SPM) regime, the low-field susceptibility of an assembly of interacting particles is expected to be of the form $\chi\approx\frac{\bar\mu^2}{3k_B(T-T_0)}$, where $\bar\mu$ is the mean magnetic moment per particle and $T_0$ is an effective temperature arising from interparticle interactions. In accordance with this equation, the reciprocal of the FC susceptibility showed a linear behaviour well above $T_M$ when it was multiplied by $M^2_S(T)/M^2_S(0)$ in order to correct the temperature dependence of $\bar \mu$. The value $T_0= -170\pm 30$ K was obtained by extrapolating this linear behaviour to the temperature axis, suggesting also the existence of strong interparticle interactions with demagnetising character. 
It is worth noting that irreversibility between the ZFC and FC susceptibilities was still present up to fields as high as 70 kOe (see Fig. \ref{Fig2a_fig}), indicating that high energy barriers were freezing the system into a glassy state. 
From all these results, it can be concluded that the broad maximum at $T_M$ in the ZFC susceptibility and the irreversibility between ZFC and FC curves originate from the blocking of the relaxation of the particle magnetisation and the freezing associated with frustrated interactions among particles. Both processes run parallel in the system giving rise to a collective state at low temperature which shares with spin-glass-like systems in bulk form most of the characteristic features of a glassy state.
%----------------------------FIG.11--------------------------------------------- 
\begin{figure}[tp] 
\centering 
\includegraphics[width= 0.7\textwidth, angle= 270]{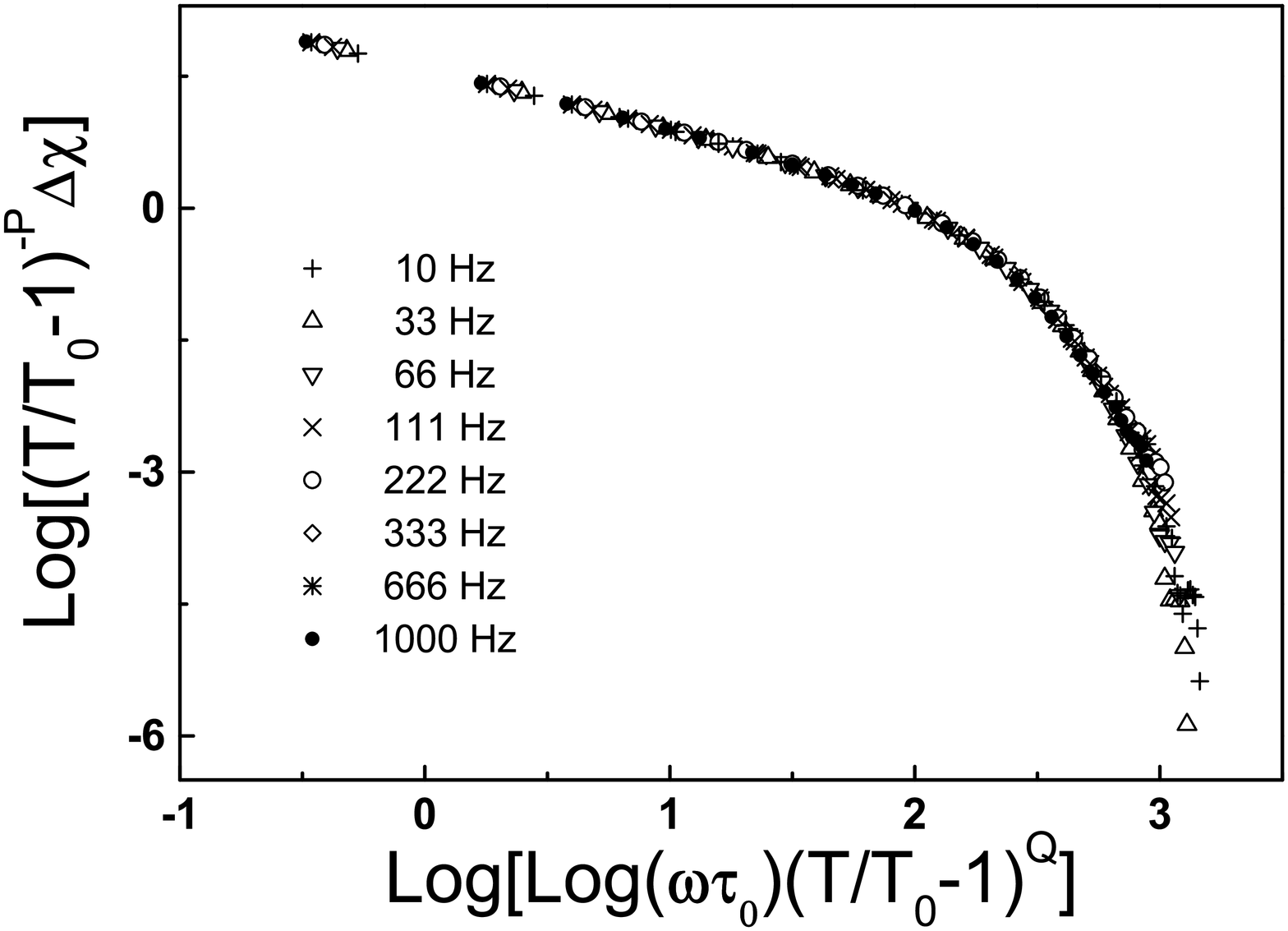}
\caption{Activated dynamic scaling for the in-phase AC
susceptibility shown in Fig. \ref{Fig3a_fig}.
}  
\label{Fig4a_fig} 
\end{figure}
%----------------------------FIG.11--------------------------------------------- 

Further insight into the dynamics of the processes responsible for the glassy state in this system can be gained by analysing the in-phase component of the AC susceptibility ($\chi_{AC}'$) measured at different frequencies from 75 to 300 K \cite{Batllejm96}. 
The measured curves (see Fig.\ref{Fig3a_fig}) behave as the DC low-field susceptibility, but the temperature at which the peak locates increases with the measuring frequency as expected for a blocking/freezing process. 
The Vogel-Fulcher law \cite{Tholencessc80} describes the slowing down of a system composed of magnetically interacting clusters as the temperature is reduced and can be expressed in the form 
\begin{eqnarray}
\label{EQ_VF_Eq}
\nu^{-1}=\tau_0 \ \exp\left[\left\langle E \right\rangle /k_B(T_M-T_0) \right] \ ,
\end{eqnarray}
where $T_0$ is an effective temperature with a similar origin to that used to reproduce the DC susceptibility in the SPM regime and $T_M$ is a characteristic temperature signalling the onset of the blocking process (e.g. the temperature of the peak position in the AC susceptibility). 
%----------------------------FIG.12--------------------------------------------- 
\begin{figure}[tp] 
\centering 
\includegraphics[width= 1.0\textwidth]{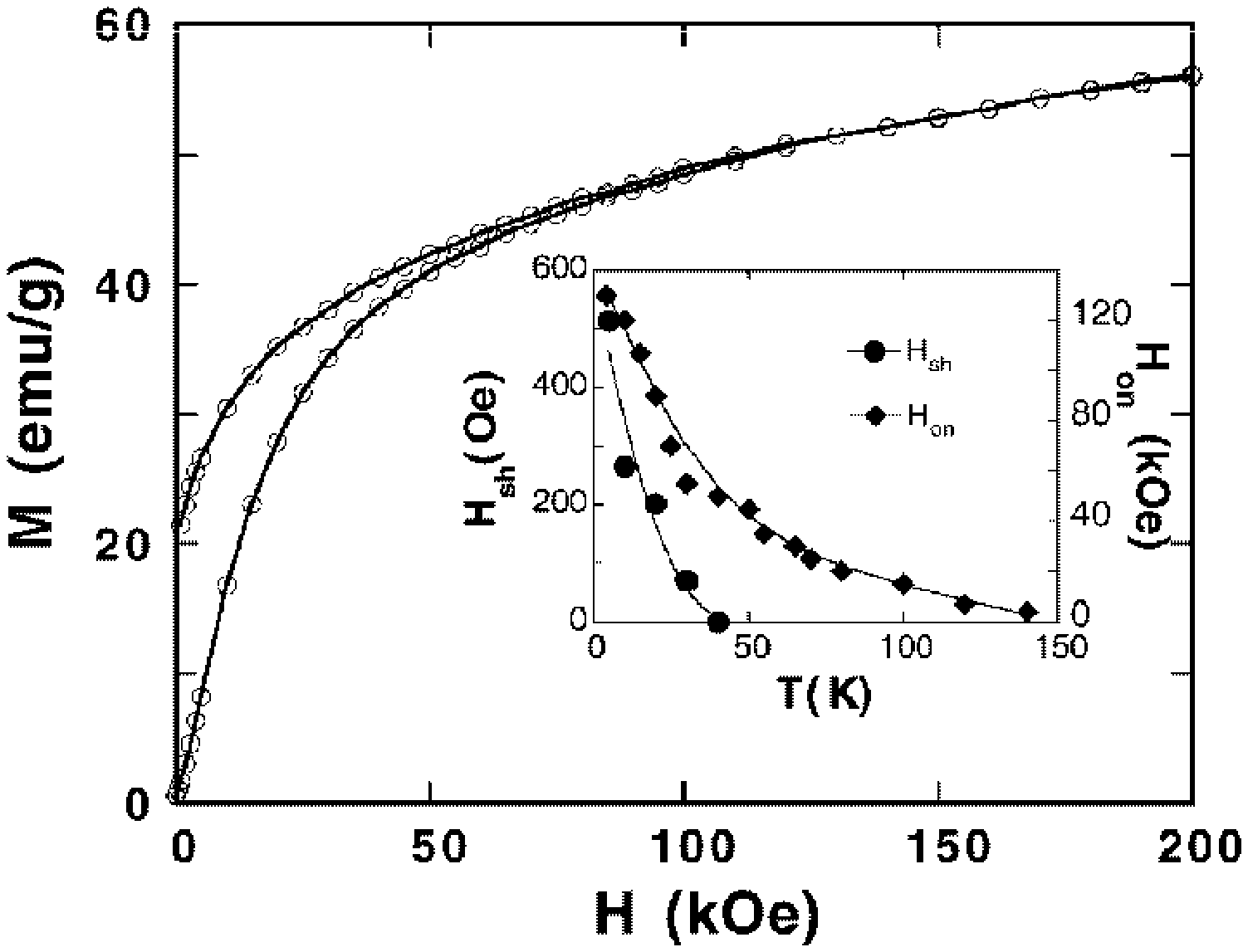}
\caption{Magnetisation vs magnetic field at 5 K for a nanocrystalline 
BaFe$_{10.4}$Co$_{0.8}$Ti$_{0.8}$O$_{19}$ sample. The inset
shows the dependence on the temperature of the onset of the
irreversibility ($H_{on}$) and the shifting of the hysteresis
($H_{sh}$) loops after an FC process at 70 kOe.
}  
\label{Fig5a_fig} 
\end{figure}
%----------------------------FIG.12--------------------------------------------
Experimental data can be fitted to Eq. \ref{EQ_VF_Eq} as shown in the inset of Fig. \ref{Fig3a_fig}, giving the following values for the parameters: 
$\tau_0=10^{-12}$ s, $\left\langle E \right\rangle= 2.45\times 10^{-13}$ erg and $T_0=159$ K, a value which is very close to the effective temperature deduced from the reciprocal of the FC susceptibility. 

The good agreement of the experimental data with the Vogel-Fulcher law evidences that the phenomenon taking place at $T_M$ is related to the blocking of an assembly of interacting particles rather than a collective freezing as that occurring in a canonical spin-glass system. As a consequence, the temperature dependence of the in-phase AC susceptibility at different frequencies can be scaled within the framework of a model based on the activated dynamics \cite{Malozemoffprb86,Fisherjap87}, in which it is supposed that $\Delta\chi=t^P G\left[-t^Q\log(\omega\tau_0) \right]$,
with $\Delta\chi=(\chi_0-\chi'_{AC})/\chi_0$, where $\chi_0$ is the equilibrium susceptibility at zero frequency, $P$ and $Q$ are the critical exponents, $t=T/T_0-1$ is the reduced temperature, and $G$ is the scaling function. 
The quality of the achieved scaling is shown in Fig. \ref{Fig4a_fig}, in which the values of the parameters are as follows: $P=0.45\pm0.2$, $Q=0.7\pm0.1$, $T_0=150\pm15$ K and $\tau_0=10^{-12}$ s. 
On the contrary, AC susceptibility cannot be scaled assuming a critical slowing down with the form of a power law \cite{Hohenbergrmp77}, fact which almost discards the occurrence of a true spin-glass transition in this assembly of interacting nanoparticles.

More evidences of the glassy behaviour are obtained from the study of the isothermal magnetisation as a function of the magnetic field. The hysteresis loop at 5 K in the first quadrant with a maximum applied field of 200 kOe is shown in Fig. \ref{Fig5a_fig}, displaying the typical features of fine-particle systems: the saturation magnetisation is about half of the bulk value, the high-field differential susceptibility is about double and the coercive field is about four times larger. Moreover, the onset of irreversibility takes place at 125 kOe, which is much larger than the typical values for bulk barium ferrites (thousands of Oe), and the hysteresis loop recorded after FC the sample at 70 kOe is shifted ca. 500 Oe in the opposite direction to the cooling field (magnetic training effect). The thermal dependence of the onset of the irreversibility and the shifting of the hysteresis loops are shown in the inset of Fig. \ref{Fig5a_fig}. The latter decays rapidly to zero as the temperature is raised and glassy behaviour disappears, while the former decreases slowly towards a value larger than the intrinsic anisotropy field because of the surface anisotropy contribution.

\begin{figure}[tp] 
\centering 
\includegraphics[width= 1.0\textwidth]{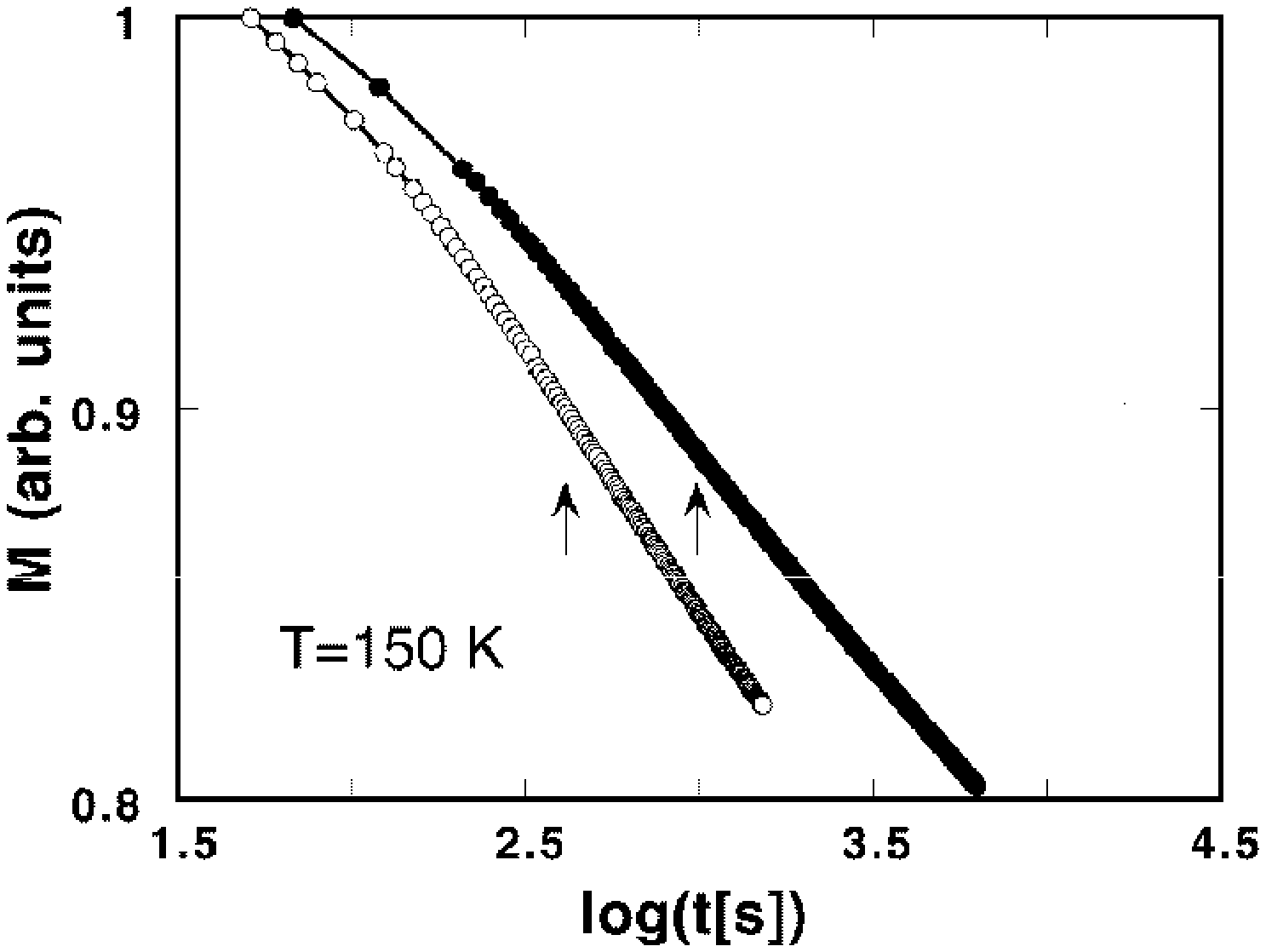}
\caption{Relaxation curves of a nanocrystalline 
BaFe$_{10.4}$Co$_{0.8}$Ti$_{0.8}$O$_{19}$ sample at 150 K after an FC process at
200 Oe for waiting times of $5\times10^2$ s (open circles) and $10^3$ s
(solid circles). Arrows indicate the position of the
inflection point of the curves, which is located at about the
waiting time.
}  
\label{Fig6a_fig} 
\end{figure}

Ageing is also commonly considered a characteristic feature of systems with enough magnetic frustration as to give rise to a multivalley energy structure at low temperature. In particular, it has been observed in fine-particle systems with strong interactions \cite{Jonssonprl95,Djurbergprl97,Mamiyaprl99} and in many glassy systems in bulk form, e.g. canonical spin glasses and dilute antiferromagnets \cite{Lundgreenprl83,Labartaprb91}. 
The phenomenon consists in the slow evolution of the magnetic order at a microscopic scale while an external magnetic field is applied, without changing significantly the net magnetisation of the system. The waiting time elapsed before the removal of the applied magnetic field is a relevant parameter determining the shape of the subsequent relaxation curves. The system studied in this work also showed ageing below about the temperature of the maximum of the ZFC curve. Fig. \ref{Fig6a_fig} shows the relaxation curves at 150 K after an FC process from room temperature at 200 Oe, for waiting times of $5\times 10^2$ and $10^3$ s. 
The characteristic trends of ageing were present in these curves proving the existence of a glassy state: (i) the relaxation rate decreased as the waiting time increased, and (ii) the relaxation curves plotted on a logarithmic time scale showed an inflection point at an elapsed time which was roughly the waiting time.

All these phenomena are common to many ferrimagnetic and antiferromagnetic oxides in the form of fine particles \cite{Kodamaprb99,Kodamaprl96,Martinezprl98,Montseprb99,Kodamaprl97,Moralescm99,Troncjm03,Dormannjm98,
Troncjm00}, suggesting that they are part of the characteristic
fingerprints of the glassy behaviour associated with the collective state existing at low temperature for strong interacting particles and/or of the surface effects appearing in the nanoscale limit.
%----------------------------FIG.13--------------------------------------------- 
%----------------------------FIG.13--------------------------------------------- 

\subsection{Surface to core exchange coupling and interparticle interactions}

Surface effects dominate the magnetic response of fine particles since for diameters smaller than ca. 2 nm more than one third of the total spins are located at the surface. Consequently, the ideal picture of a superspin formed by the collinear arrangement of all the spins of the particle is no longer valid, and the misalignment of the surface spins yields strong deviations from the bulk behaviour. 
This is also true for particles of many ferrimagnetic oxides with strong exchange interactions, in which magnetically competing sublattices usually exist. In these cases, broken bonds and defects at the surface layer destabilise magnetic order giving rise to magnetic frustration, which is enhanced with the strength of the magnetic interactions. As a consequence, the profile of the magnetisation is not uniform across the particle, the surface layer being more demagnetised than the core spins. In early models, this fact was modelled postulating the existence of a dead magnetic layer giving no contribution to the magnetisation of the particle \cite{Berkowitzjap68}. Coey proposed the existence of a random canting of the surface spins due to the competing antiferromagnetic interactions between sublattices to account for the reduction of the saturation magnetisation observed in $\gamma$-Fe$_2$O$_3$ ferrimagnetic nanoparticles \cite{Coeyprl71}. He found that a magnetic field of 5 T was not enough to achieve full alignment of the spins in the field direction for particles 6 nm in size. A further verification of the existence of spin canting in nanoparticles of different ferrimagnetic oxides ($\gamma$-Fe$_2$O$_3$, NiFe$_2$O$_4$, CoFe$_2$O$_4$, CuFe$_2$O$_4$) was gained by using M\"ossbauer spectroscopy \cite{Coeyprl71,Hanedajap82}, polarised \cite{Linjm95} and inelastic neutron scattering \cite{Gazeaueul97}, and ferromagnetic resonance \cite{Gazeaujm98}. However, in contrast with the original suggestion by Coey that spin canting occurs chiefly at a surface layer due to magnetic frustration, some works based on M\"ossbauer spectroscopy support the idea that it is a finite-size effect which is uniform throughout the particle \cite{Parkerprb93,Pankhurstprl91,Moralesjpcm97,Linderothjap94,Kodamaprl97}. In fact, the origin of the spin misalignment observed in ferrimagnetic nanoparticles is still under discussion and there is not a complete understanding of the phenomenon. Nevertheless, spin canting has not been found in metallic ferromagnetic particles, which reinforces the hypothesis that magnetic frustration originated from antiferromagnetically competing sublattices is a necessary ingredient to explain the non-collinear arrangement of the particle spins.

Based on all the foregoing results and on the glassy properties observed at low temperatures, a model has been proposed \cite{Kodamaprl97,Martinezprl98} suggesting the existence of a magnetically ordered core surrounded by a surface layer of canted spins, which undergoes a spin-glass-like transition to a frozen state below a certain temperature, $T_f$. 
The glassy state of the surface below $T_f$ creates an exchange field acting over the ordered core of the particle, which could be responsible for the shifting of the hysteresis loops after an FC process by a mechanism similar to that giving rise to exchange bias in layered structures \cite{Noguesjm99,Stampsjpd00}. 
Besides, the magnetic frustration at the surface may increase the effective anisotropy energy  giving rise to both the high-field irreversibility observed in the ZFC-FC curves and the high values of the closure field in the hysteresis loops. In fact, these anomalous and enhanced properties vanish or are strongly reduced above $T_f$, a fact that indicates that they are related to some kind of frozen state occurring below $T_f$.

The existence of a frozen state at the particle surface has been experimentally established by different techniques \cite{Kodamaprl96,Montseprb99,Kodamaprb99,Martinezprl98,Kodamajap97,Koksharovprb00}. For instance, electron paramagnetic resonance measurements in iron-oxide nanoparticles 2.5 nm in diameter diluted in a polyethylene matrix \cite{Koksharovprb00}, showed a sharp line broadening and a shifting of the resonance field on sample cooling below $T_f$, which is about the temperature at which an anomaly in the FC susceptibility was also observed. 
Furthermore, the study of the field dependence of $T_f$ in $\gamma$-Fe$_2$O$_3$ nanoparticles of 10 nm  average size demonstrated that this magnitude follows the Almeida-Thouless line  \cite{Martinezprl98} which is a characteristic behaviour found in many magnetic glassy phases. This field dependence of $T_f$ was not affected by diluting the magnetic particles with a nanometric SiO$_2$ powder \cite{Martinezprl98}, which excludes that the glassy state in these systems could originate from interparticle interactions only.

A micromagnetic model at atomic scale proposed by Kodama and Berkowitz \cite{Kodamaprl96,Kodamaprb99} and several numerical simulations of a single ferrimagnetic particle with a variety of assumptions but with the common condition of free boundaries at the surface 
\cite{Kachkachiepj00,Dimitrovprb94,Dimitrovprb95,Kachkachijap02,Kachkachiprb02,Labayjap02,Trohidouprb90,Kechrakosjm98,Trohidoujap97,Trohidoujap98,Trohidoujap99},
have evidenced the non-uniformity of the magnetisation profile across the ferrimagnetic particle, with a fast decreasing towards the surface. 
These results demonstrate that the non-uniform profile of the particle magnetisation is merely a surface effect. 
However, enhanced anisotropy (normal to the surface) \cite{Neeljpr54}, vacancies and broken bonds at the surface have to be included in these particle models in order to induce enough magnetic frustration so as to freeze the disordered surface layer giving rise to a glassy state \cite{Kodamaprb99,Kachkachiepj00}. 
Therefore, surface and finite-size effects seem not to be enough as to produce the glassy layer, even in the case of ferrimagnetic particles with competing antiferromagnetic sublattices, surface anisotropy and disorder being necessary additional ingredients.

In highly concentrated samples (e.g. powder samples) with a random distribution of easy axis, interparticle interactions are a supplementary source of magnetic frustration which may lead to a frozen collective state of the particle spins at low temperature, apart from the effects of the surface-to-core exchange coupling discussed above. 
In fact, both processes may occur in parallel contributing simultaneously to the glassy phenomenology in ferrimagnetic fine particles. The main types of magnetic interactions that can be found in fine-particle assemblies are dipole-dipole interaction, which always exists, and exchange interactions through the surface of the particles being in close contact \cite{Altbirprb96}. 
Taking into account the anisotropic character of dipolar interactions, which may favour parallel or antiparallel arrangements of the spins depending on the geometry, and the random distribution of local easy axis, concentrated samples have the required elements of magnetic frustration to give rise to a glassy state. The complex interplay between both sources of magnetic disorder determines the state of the system and its dynamical properties. In particular, the effective distribution of energy barriers, which block the inversion of particle spins, is further enhanced by magnetic frustration leading to high field irreversibilites. 

In the limit of strong dipolar interactions, individual energy barriers coming from intrinsic anisotropy of the particles can no longer be considered, the total energy of the assembly being the only relevant magnitude. In this limit, relaxation is governed by the evolution through an energy landscape similar to that of a spin glass with a complex hierarchy of energy minima. 
The inversion of one particle moment may also modify the energy barriers of the whole assembly. 
Consequently, the energy barrier distribution may evolve as the total magnetisation of the system relaxes \cite{Berkovjm92b,Berkovprb96,OscarThesis,Ribasjap96}
and non-equilibrium dynamics may appear, for instance, showing ageing effects as observed in several fine-particle systems with strong interactions 
\cite{Djurbergprl97,Dormannjm98,Jonssonprb00,Jonssonprl95,Jonssonprb98,Mamiyaprl99}.
In this way, the occurrence of ageing phenomena in fine particles could be considered as a clear indication of a strong interacting scenario with a phenomenology which largely mimics that corresponding to spin glasses, including memory effects in which magnetic relaxation depends on heating or cooling rates and/or on the cycles followed by the temperature \cite{Mamiyaprl99,Jonssonprb00}. However, three main differences between ageing in bulk spin-glasses  and fine particles may be established: i) the dependence of the relaxation on the waiting time is weaker (see, for instance, Fig. \ref{Fig6a_fig}); ii) in the collective state, the relaxation times are widely distributed and strongly dependent on temperature; and iii) the moment of the largest particles is blocked during long time scales, acting as a random field throughout the system.

Finally, one of the facts that complicates the study of these systems is the coexistence of the freezing associated with magnetic frustration and the intrinsic blocking of the particles. Consequently, depending on the time window of the experiment, one or both phenomena are observed. For example, blocking processes usually determine the results of M\"ossbauer spectroscopy, since the measured blocking temperature decreases with increasing interactions \cite{Morupprl94}, while freezing phenomena determine the temperature location of the cusp of the real part of the AC susceptibility for concentrated samples, which moves to higher temperatures with increasing interactions \cite{Jonssonprb98,Dormannjpc88,Dormannjm98}.
%----------------------------FIG.14--------------------------------------------- 
\begin{figure}[tp] 
\centering 
\includegraphics[width= 1.0\textwidth]{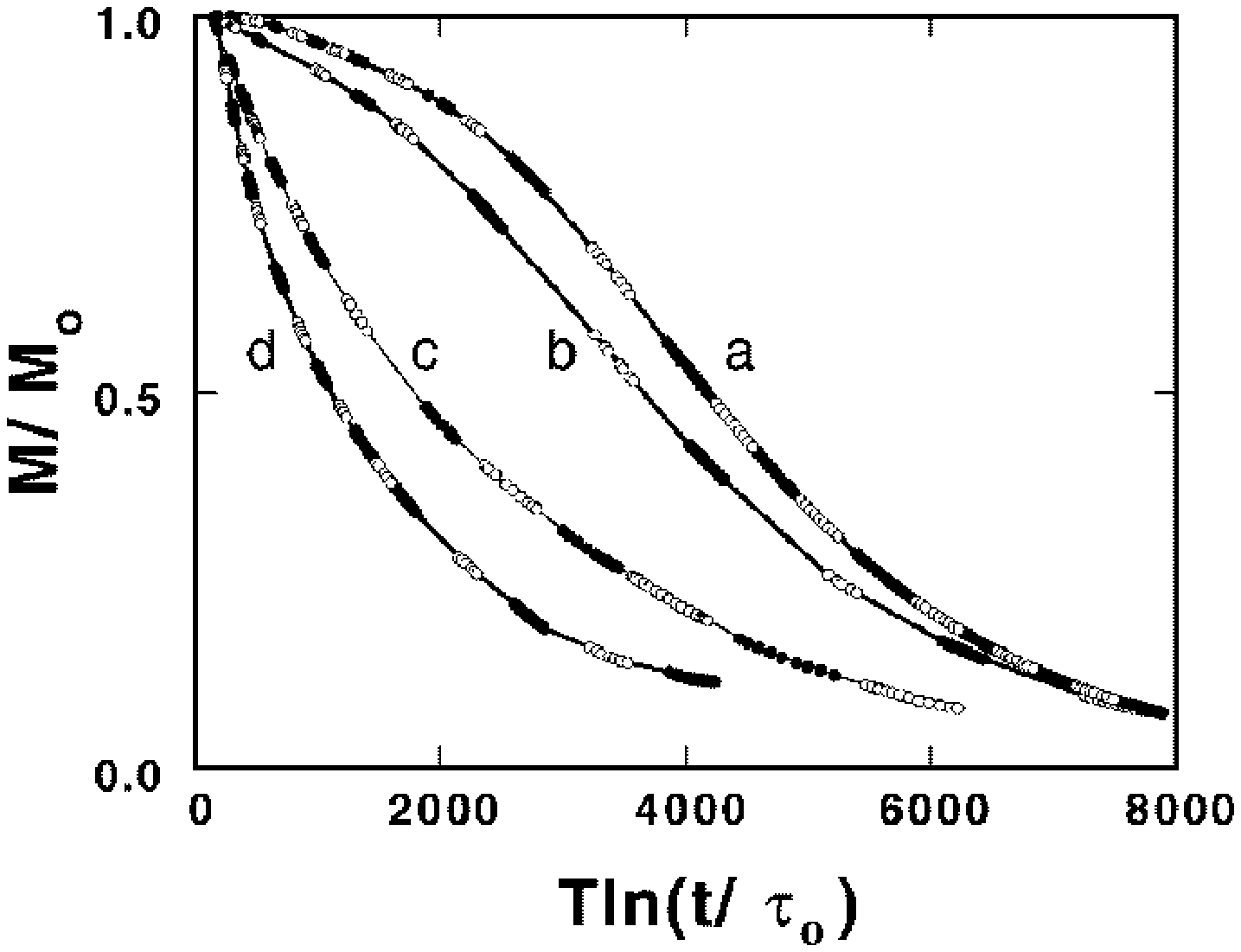}
\caption{$T\ln(t/\tau_0)$ scaling of the magnetic relaxation
measured in the temperature range 5-230 K after an FC process
at 200 Oe (a), 500 Oe (b), 10 kOe (c), 50 kOe (d). Alternated open and full circlesb correspond to consecutive different temperatures.
}  
\label{Fig7a_fig} 
\end{figure}
%----------------------------FIG.14--------------------------------------------- 

\subsection{Effective distribution of energy barriers from relaxation measurements}

One way to obtain a further insight into the nature of the glassy phenomenology described in the preceding sections consists in analysing the time dependence of the magnetisation of the system in terms of the so-called $T\ln(t/\tau_0)$ scaling \cite{Labartaprb93,Prejeanjpe80,Omarijpe84}, since this method allows one to calculate the effective distribution of energy barriers \cite{Iglesiasjm95,Iglesiaszpb96,Balcellsprb97}. 
In this scaling procedure, the value of $\tau_0$ (characteristic attempt time) is chosen so as to make all the relaxation curves, measured at different temperatures, scale onto a single master curve, which stands for the relaxation curve at the lowest measuring temperature extended to very long times. The effective distribution of energy barriers is then obtained by performing the derivative of the experimental master curve with respect to the  $T\ln(t/\tau_0)$ variable \cite{Iglesiasjm95,Iglesiaszpb96,Balcellsprb97}. 
%----------------------------FIG.15--------------------------------------------- 
\begin{figure}[tp] 
\centering 
\includegraphics[width= 1.0\textwidth]{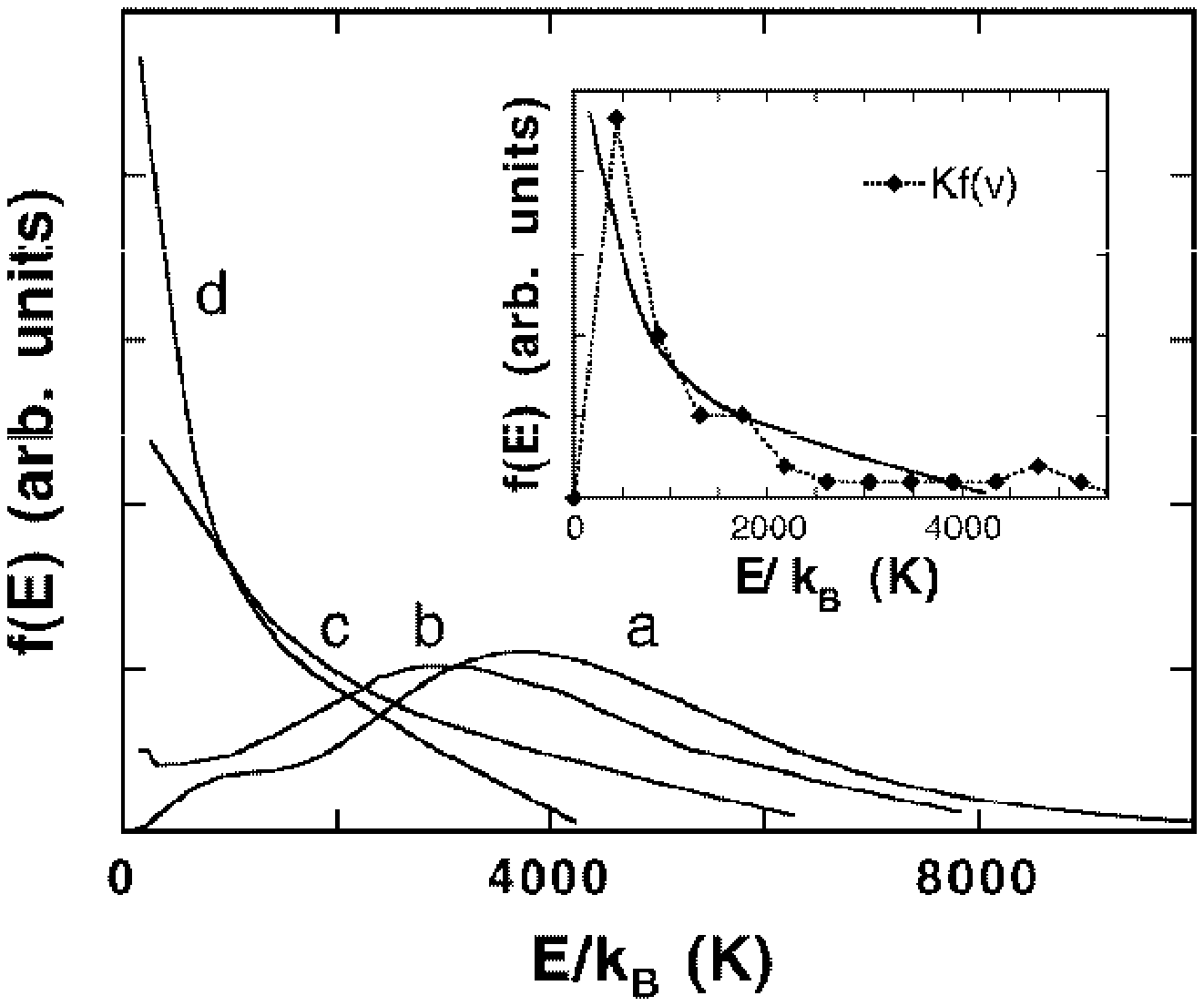}
\caption{Effective distribution of energy barriers
obtained from the scaling of Fig.\ref{Fig7a_fig}. The labels of the
curves are as in Fig.\ref{Fig7a_fig}. Inset: energy distribution obtained after an FC process at 50 kOe [curve (d)] compared with $Kf(v)$ (symbols and dashed line respectively). $K$ stands for the anisotropy constant and $f(v)$ is the volume distribution.
}  
\label{Fig8a_fig} 
\end{figure}
%----------------------------FIG.15---------------------------------------------
%
However, we could wonder whether this method is applicable in the strong interacting regime, for which it is not possible to identify individual energy barriers blocking the reversal of the particle moments as pointed out before. Monte Carlo simulations and experimental results \cite{OscarThesis,Ribasjap96,Iglesiasjm95,Batlleprb97} have demonstrated that a good $T\ln(t/\tau_0)$ scaling is also achievable in this case, from which a static effective distribution of energy barriers (non-time-evolving) may also be derived. This energy distribution is similar to that observed in non-interacting systems, even though it is not only a direct consequence of the distribution of particle anisotropies, being an average of the contributions due to volume, surface and shape anisotropies, and interparticle interactions. Through this simple method, it is possible to ascertain the differences between fine-particle systems with glassy phenomenology, for which the existence of a large number of quasidegenerate energy levels makes the dynamics complex, since their actual energy barrier distribution is time-dependent.
 
As an example of this procedure applied to a strong interacting system, in Fig. \ref{Fig7a_fig}, we show the scaling of the relaxation data measured after FC of the sample at different fields for the BaFe$_{10.4}$Co$_{0.8}$Ti$_{0.8}$O$_{19}$ particles discussed in Section \ref{Anomalous_Sec}. The value of $\tau_0$ used in this scaling is $10^{-12}$ s, which is consistent with those values deduced from the frequency dependence of the maximum of the real part of the AC susceptibility (Vogel-Fulcher law) and the activated dynamic scaling of these data (see Figs. \ref{Fig3a_fig}, \ref{Fig4a_fig}). 
It is evident from Fig. \ref{Fig7a_fig} that the cooling field drastically modifies the relaxation curves, which demonstrates that the initial arrangement of the particle moments (FC state) determines the time evolution of the magnetisation when interparticle interactions are strong \cite{GarciaOteroprl00,Portoepjb02,Ulrichprb03}. This is in contrast with the noninteracting case, for which the observed results are independent of the cooling field, at least for fields lower than that at which the lowest energy barriers start to be destroyed \cite{Iglesiasjap02,Iglesiascostp3}. 

\subsection{Effects of the magnetic field on the glassy state}
The effective distribution of energy barriers, $f(E)$, characterising the glassy state of the assembly as a function of the cooling field, can be obtained from the scaling curves in Fig. \ref{Fig7a_fig} by the procedure detailed in Refs. \cite{OscarThesis,Labartaprb93,Balcellsprb97,Iglesiasjap02,Iglesiascostp3}, the results being shown in Fig. \ref{Fig8a_fig}. At low cooling fields, $f(E)$ extends to extremely high energies, and the energy corresponding to the maximum of the distribution is one order of magnitude higher than that expected from bulk anisotropy, $Kf(V)$, where $K$ is the bulk anisotropy constant \cite{Batllejap91} and $f(V)$ is the volume distribution derived from TEM \cite{Gornertie94}. 
In contrast, $f(E)$ progressively resembles $Kf(V)$ as the cooling field increases (see inset of Fig. \ref{Fig8a_fig}). 
This field dependence of $f(E)$ may be interpreted as follows. The component of the energy barrier distribution centred at high energies, which is dominant at low cooling fields, is then attributed to the collective behaviour associated with the glassy state, since particle magnetisation is mostly randomly distributed in the FC state. Nevertheless, $f(E)$ at high cooling fields, which is centred at much lower energies, corresponds to the intrinsic anisotropy of the individual particles. In the high-field-cooled state, the particle magnetisation is mostly aligned parallel to the field. Therefore, the overall dipolar interactions are demagnetising and their effect can be considered through a mean field which reduces the height of the energy barrier associated with the intrinsic anisotropy of the particles \cite{OscarThesis,Iglesiasjap02}. 
%----------------------------FIG.16--------------------------------------------- 
\begin{figure}[tp] 
\centering 
\includegraphics[width= 1.0\textwidth]{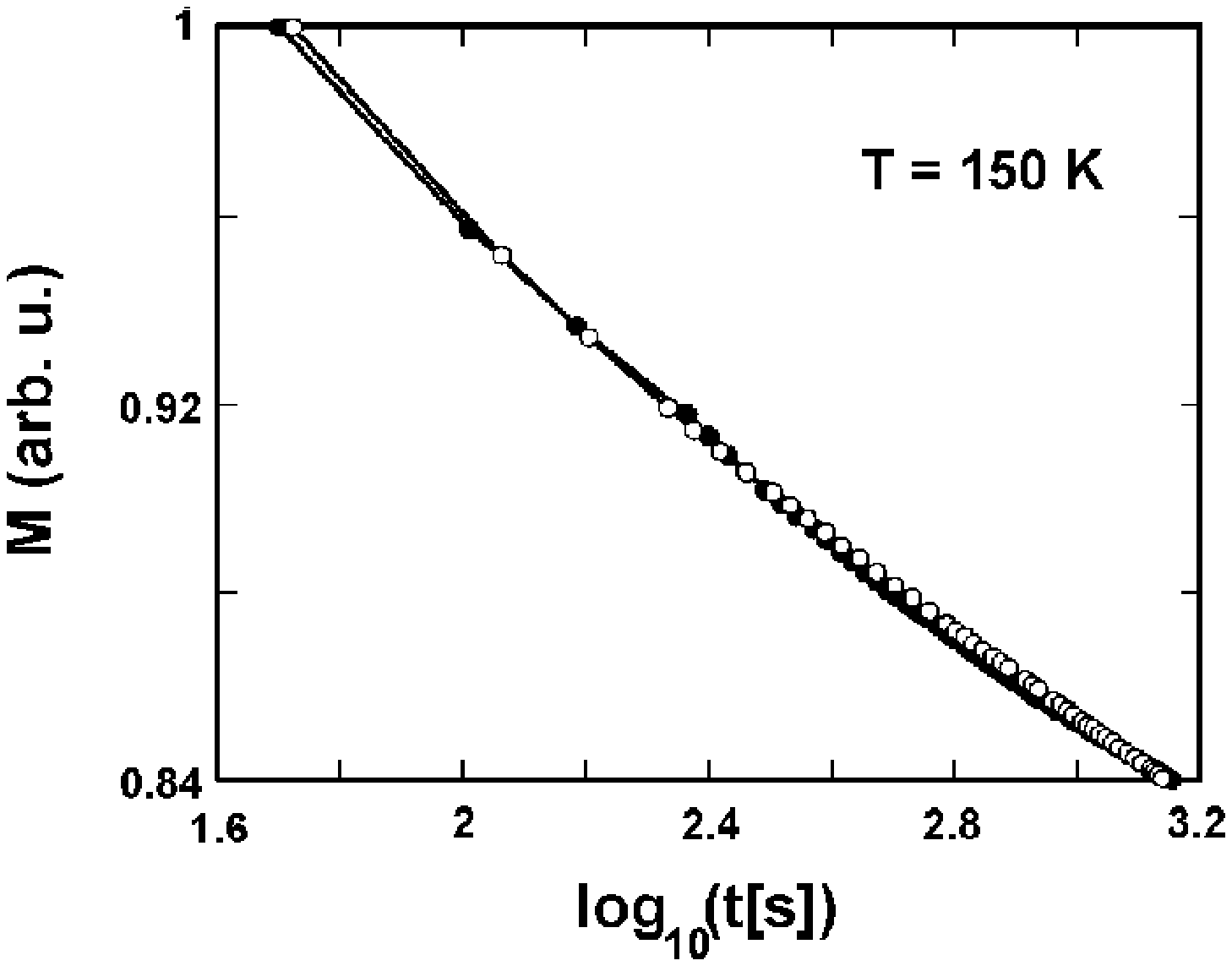}
\caption{Relaxation curves at 150 K after an FC process at
10 kOe for waiting times of $5\times 10^2$ s (open circles) and $10^3$ s
(solid circles).
}  
\label{Fig9a_fig} 
\end{figure}
%----------------------------FIG.16--------------------------------------------- 
Consequently, a slight shift to lower energies is observed in $f(E)$ with respect to $Kf(V)$ (see inset of Fig. \ref{Fig8a_fig}). At intermediate fields, a bimodal $f(E)$ arising from both contributions is observed, their relative importance being determined by the strength of the cooling field (see $f(E)$ obtained after FC processes at 200 and 500 Oe in Fig. \ref{Fig8a_fig}). This interpretation is also confirmed by repeating the ageing relaxation experiments shown in Fig. \ref{Fig6a_fig} with a cooling field of 10 kOe instead of 200 Oe. In this case, no significant differences in the relaxation curves are detected as the waiting time is varied (see Fig. \ref{Fig9a_fig}). Accordingly, the characteristic ageing associated with the glassy state can be detected only at low enough cooling fields. 
Therefore, the glassy state in particle assemblies originated from strong interparticle interactions can be erased by field-cooling the sample at moderate fields, which precludes the occurrence of a true spin-glass transition at the freezing temperature. Even though interacting particle systems seems to exhibit an effective irreversibility line similar to that found in other cluster and spin-glass systems in the bulk form, the magnetic fields at which the disordered state is erased are very low compared, for example, to either the anisotropy field of the particles or from those values corresponding to archetypal spin glasses \cite{Kenningprl91}.

All in all, cooling fields monitor the dynamics of interacting fine magnetic particles through determining the initial state of the magnetic moment arrangement. Consequently, at high cooling fields the dynamics of the system is mostly dominated by the intrinsic energy barriers of the individual particles, while at low cooling fields, the energy states arising from collective glassy behaviour play the dominant role. Thus, care should be taken when comparing relaxation data from isothermal remanent magnetisation and thermoremanent magnetisation, since the initial magnetic state may be very different in both kinds of experiments depending on the field strength. 
Similar results have been obtained by J\"{o}nson et al. \cite{Jonssonprb01} in strongly interacting FeC nanoparticles, for which the collective glassy dynamics can be destroyed by the application of moderate fields and also in the Monte Carlo simulations by Ulrich et al. \cite{Ulrichprb03}. 
Fine-particle systems are thus relevant because, although they display an important degree of magnetic frustration, the collective state may be destroyed by the application of a moderate field, which precludes a true spin-glass behaviour.
%%%%%%%%%%%%%%%%%%%%%%%%%%%%%%%%%%%%%%%%%%%%%%%%%%%%%%%%%%%%%%%%%%%%%%%%%%%%%%
\section{Monte Carlo simulations}
%Up to the moment, there is no unique model giving a clear-cut interpretation of the all the %above-mentioned peculiar phenomenology of ferrimagnetic nanoparticles. a, a number of %works studying part of the issues have been published recenty.  
%
%A number of works studying the peculiar phenomenology of ferrimagnetic nanoparticles	 have %been published.  
%
Numerical modeling and, in particular, Monte Carlo (MC) simulations, are well suited
to discern between the contributions to the glassy behavior due to inter-particle interactions and single-particle effects, such as surface and finite-size effects.  
By considering the particles as dipolar interacting superspins, MC simulations reproduce the enhancement of the magnetic viscosity and the increase in the energy barriers \cite{OscarThesis,Ribasjap96} and also  the non-equilibrium dynamics showing ageing and memory effects \cite{Anderssonprb97}. On the contrary, some relaxation simulations starting from different states discard the occurrence of cooperative freezing since all the curves converge to the equilibrium state \cite{GarciaOteroprl00,Portoepjb02} whereas on similar simulations performed in the absence of a magnetic field the system seem to approach some glassy ferromagnetic state \cite{Ulrichprb03}.
Alternatively, by taking a single particle as the simulation unit, numerical models can also be used to study intrinsic effects related to the finite volume and boundary limits of the particle, disregarding collective behaviour due to inter-particle interactions. In this approach, the atomic spins and the underlying lattice are considered in detail focusing the attention on the magnetic characterisation in terms of the microscopic structure. In what follows, we summarise the chief previous studies of single particle models.

The first atomic-scale model of the magnetic behaviour of individual ferrimagnetic 
nanoparticles is due to Kodama and Berkowitz \cite{Kodamaprb99}. The authors presented results of calculations of a micromagnetic model of maghemite particles which were based on an energy minimisation procedure, instead of the Monte Carlo method. They used Heisenberg spins with enhanced anisotropy at the surface with respect to the core and included vacancies and broken bonds at the surface, arguing that these are indeed necessary to obtain hysteresis loops with enhanced coercivity and high-field irreversibility. 
Later, Kachkachi et al. \cite{Kachkachiepj00,Kachkachijm00,Kachkachiphya01}  
performed MC simulations of a maghemite particle described by a 
Heisenberg model, including exchange and dipolar interactions and 
using surface exchange and anisotropy constants different to those of the bulk. 
Their study was mainly focused on the thermal variation of the surface 
(for them consisting of a shell of constant thickness) and core magnetisation, 
concluding that surface anisotropy is responsible for the non-saturation of the
magnetisation at low temperatures. More recently \cite{Kachkachijap02,Kachkachiprb02}, they studied the influence of surface anisotropy in the zero-temperature hysteretic properties of a ferromagnetic particle by means of numerical evaluation of Landau-Lifschitz equations.

Other computer simulations studying finite-size and surface effects on ferro- 
and antiferromagnetic cubic lattices have also been published.
Bucher and Bloomfield \cite{Bucherprb92} presented a quantum mechanical calculation and performed MC simulations of a Heisenberg model to explain the measured reduction in the magnetic moment of small free Co clusters.
Trohidou et al. \cite{Trohidouprb90,Trohidoujap99} performed MC simulations of 
AF small spherical clusters. By using an Ising model on a cubic lattice 
\cite{Trohidouprb90}, they computed the thermal and magnetic field dependence of 
the magnetisation and structure factor, concluding that the particle behaved as a 
hollow magnetic shell. By means of a Heisenberg model \cite{Trohidoujap99}
with enhanced surface anisotropy, they studied the influence of different kinds 
of surface anisotropy on the magnetisation reversal mechanisms 
and on the temperature dependence of the switching field. 
Dimitrov and Wysin \cite{Dimitrovprb94,Dimitrovprb95} studied the 
hysteresis phenomena of very small spherical and cubic ferromagnetic (FM) fcc clusters of 
Heisenberg spins by solving the Landau-Lifshitz equations. 
They observed an increase of the coercivity 
with decreasing cluster size and steps 
in the loops due to the reversal of surface spins at different fields.
However, they did not consider the finite temperature effects.
Also Altbir and co-workers \cite{Altbirprb01} investigated the evolution of nanosize Co cluster under different thermalisation processes by MC simulations.
%Pioneering studies of the critical properties of systems with free boundaries were 
%initiated by Binder and Landau \cite{Binderprl84,Binderdomb83,Landauprb90} 
%back in the 80s, giving scaling laws and exponents for the surface 
%magnetisation and relating them to the results for the bulk. 
%More recently, Pleimling et al. \cite{Pleimlingeur98} and 
%Zhao et al. \cite{Zhaoprb00} also have studied by MC simulation the 
%effect of different kinds of surface roughness and imperfections on 
%the magnetisation processes of ferromagnetic films and taking into account 
%exchange constants at the surface different from those of the bulk. 
%Their main observation was a reduction of magnetisation due to the 
%reduced coordination at the surface.

In what follows, we will present the results of extensive MC simulations 
\cite{Iglesiasprb01,Iglesiasjap01,Iglesiasrodas01}
which aim at clarifying what is the specific role of the finite size and surface on the magnetic properties of the particle, disregarding the interparticle interactions. 
In particular, we study the magnetic properties under a magnetic 
field and at finite temperature, thus extending other simulation works.
In choosing the model, we have tried to capture the main features of
real particles with the minimum ingredients allowing us to interpret
the results without any other blurring effects.
%-------------------------------------------------------------------------------------
\subsection{One-particle model for $\gamma$-Fe$_2$O$_3$}

In maghemite \magh with the spinel structure described in Sec. \ref{Frustration_Sec}, 
the spins interact via antiferromagnetic (AF) exchange interactions with the 
nearest neighbours on both sublattices and with an external magnetic field $H$,  
the corresponding Hamiltonian of the model being
\bea 
{\cal H}/k_{B}= 
-\sum_{\alpha,\beta=\,T,O}\sum_{i,n=1} 
          J_{\alpha\beta} S_i^{\alpha} S_{i+n}^{\beta}  
          -h\sum_{\alpha= T,O}\sum_{i=1}^{N_\alpha} S_i^{\alpha}\ ,
\eea 
where we have defined the field in temperature units as $h=\frac{\mu H}{k_B}$,
withbb $S$ and $\mu$ the spin value and magnetic moment of the Fe$^{3+}$ ion,
respectively. 
In our model, the Fe$^{3+}$ magnetic ions are represented by Ising spins 
$S_i^{\alpha}=\pm 1$, which allows us to reproduce a case with strong uniaxial 
anisotropy while keeping computational efforts within reasonable limits.
The maghemite values of the nearest neighbour exchange constants given in Sec. \ref{Frustration_Sec} are considered. 
%----------------------------FIG.1--------------------------------------------- 
\begin{figure}[t] 
\centering 
\includegraphics[width= 1.0\textwidth]{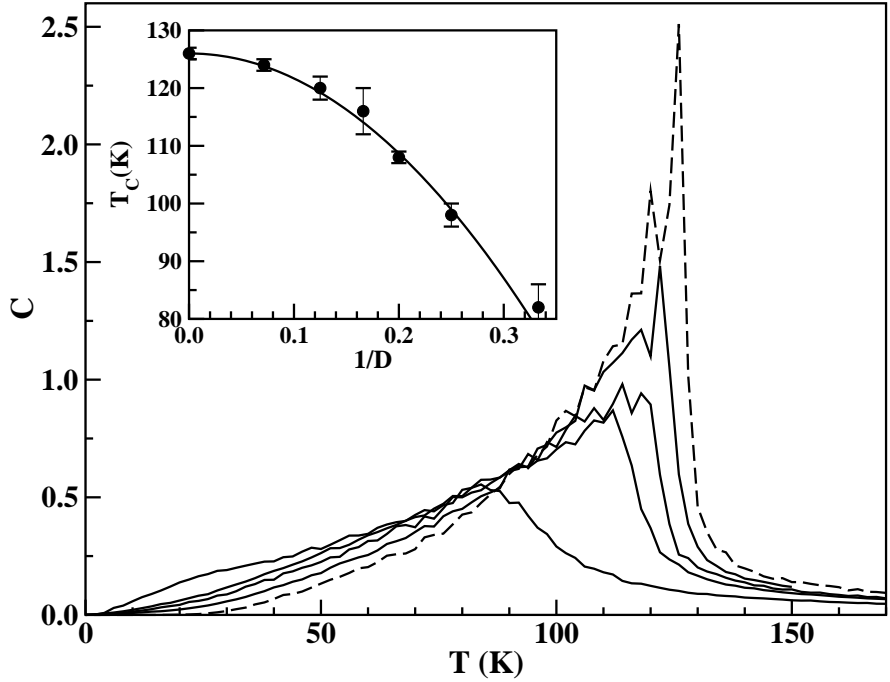}
\caption{Thermal dependence of the specific heat $C$ for different diameters $D= 3, 6, 8,  
14$ (from the uppermost curve) and periodic boundary conditions for $N= 14$ (dashed curve). Inset: Particle size dependence of the transition temperature $T_c$ from paramagnetic to ferrimagnetic phases for spherical particles with FB. 
The displayed values have been obtained from the maximum in the  
specific heat. The continuous line is a fit to Eq. \ref{TcEq}.
}  
\label{Ce_fig} 
\end{figure}
%----------------------------FIG.1--------------------------------------------- 

We have assumed two kinds of boundary conditions on the lattice defined above with the aim to study the changes induced by the finite size of the nanoparticle in the magnetic properties with respect to the bulk material. By using periodic boundary (PB) conditions for systems with large enough linear size $N$ (linear sizes are measured in multiples of the unit cell length), we will simulate the bulk behaviour. When  studying a finite-sized particle, we will cut the lattice in the shape of a sphere of diameter $D$ (measured in multiples of the unit cell length) with free boundary (FB) conditions. Moreover, in the last case, we distinguish two regions in the particle: the surface formed by the outermost unit cells, and a core formed by a sphere of spins with diameter $D_{Core}$ unit cells. The particle sizes considered range from $D= 3$ to $D=10$, corresponding to real diameters from $25$ to $83$ {\rmfamily\AA}.
The different measured magnetisations $M_{Unc}$ are given in normalised units with respect to the number of uncompensated spins, $M_{Unc}=(N_O - N_T)/N_{Total}$ the ratio of the difference of O and T spins to the total number of spins $N_{Total}$, that for an infinite lattice with PB conditions is $1/3$.

\subsection{Equilibrium properties: surface and core magnetisations}
Let us begin by studying the finite-size and surface effects on the equilibrium magnetic properties. The simulation protocol uses the standard Metropolis algorithm and proceeds in the following way.
Starting from a disordered configuration of spins at a high temperature ($T= 200$ K), the temperature is progressively reduced in steps $\delta T= -2$K. At each temperature, thermal averages of the thermodynamic quantities were performed during $10 000$ to $50 000$ MC steps after discarding the first $1000$ MC steps for thermalisation.The quantities monitored during the simulation are the energy, specific heat, susceptibility, and different magnetisations: sublattice magnetisations ($M_O, M_T$), surface, core and total magnetisation ($M_{Surf}, M_{Core}$, $M_{Total}$). Note that with the above-mentioned normalisation, $M_{Total}$ is $1$ for ferromagnetic order, $0$ for a disordered system and $1/3$ for ferrimagnetic order of the O and T sublattices.
%-------------------------------------------------------------------------------------

In Fig. \ref{Ce_fig}, we present the thermal dependence of the specific heat, $C$, for different particle diameters and we compare it to the PB case for $N=14$. In all the cases, the sharp peak in $C$ at the critical ordering temperature $T_C(D)$ is a sign of a second order transition from paramagnetic to ferrimagnetic order.
As we can see, finite-size effects are clearly noticeable in the FB case even for $D$'s as large as $D= 14$. In particular, $T_C(D)$ increases as the particle size is increased, approaching the infinite-size limit which can be estimated from the PB curve for $N= 14$ as $T_c(\infty)=126\pm 1$ K, as shown in the Inset of Fig. \ref{Ce_fig}. Finite-size scaling theory \cite{Binderprb74,Landauprb76,Barber} predicts the following scaling law for $T_c(D)$
\begin{equation} 
\label{TcEq}
\frac{T_{c}(\infty)-T_c (D)}{T_{c}(\infty)} = \left(\frac{D}{D_0}\right) ^{-1/\nu} \ .
\end{equation} 
%----------------------------FIG.2--------------------------------------------- 
\begin{figure}[tp] 
\centering 
\includegraphics[width= 1.0\textwidth]{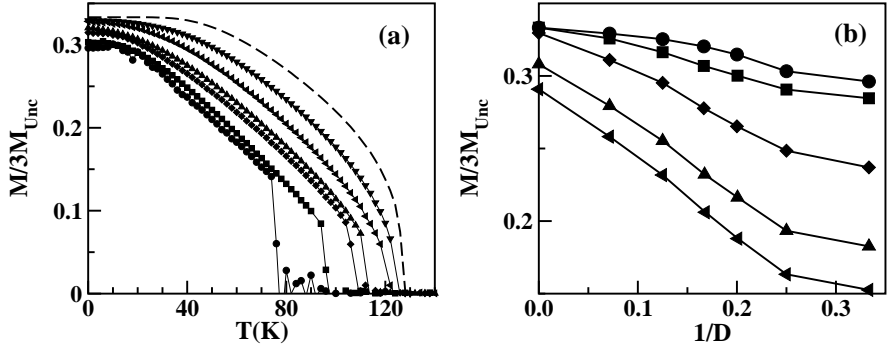}
\caption{
(a) Thermal dependence of the magnetisation $M$. The results for particle diameters $D= 3, 4, 5, 6, 8, 14$ (from the lowermost curve in circles) and PB conditions  
$N= 14$ (dashed line) are shown. 
(b) Size dependence of the magnetisation of a spherical particle at 
different temperatures $T=$ 0, 20, 40, 60, 70 K (from upper to lowermost
curves). $M_{\rm Unc}$ is the ratio of the difference of O and T spins to the total number 
of spins.  
}  
\label{M(T)_Db_fig} 
\end{figure}
%----------------------------FIG.2--------------------------------------------- 
This expression fits nicely our MC data with $D_0 = 1.86\pm 0.03$ being a microscopic length scale (in this case, it is roughly twice the cell parameter), and a critical exponent $\nu=0.49\pm 0.03$, which seems to indicate a mean field behaviour \cite{Binderprb72,Stanley}. This result can be 
ascribed to the high coordination of the O and T sublattices.
The fitted curve is drawn in Fig. \ref{Ce_fig} where deviations from 
scaling are appreciable for the smallest diameters for which corrections to 
the finite-size scaling in Eq. \ref{TcEq} may be important \cite{Landauprb76}. 
Thus, these results discard any important surface effect on the ordering temperature and are consistent with spin-wave calculations \cite{Hendriksenprb93} and old MC simulations \cite{Binderjpcs70}. 
Similar finite-size effects have been found in fine particles \cite{Tangprl91} of MnFe$_2$O$_4$, but with a surprising increase of $T_c(D)$ as $D$ decreases, which has been attributed to surface effects due to the interactions with the particle coating.
%-------------------------------------------------------------------------------------

In order to better understand how the finite-size effects influence the magnetic order it is also interesting to look at the thermal dependence of the magnetisation for different particle sizes. 
In Fig. \ref{M(T)_Db_fig}a, we compare the results for spherical particles with different $D$ to the bulk behaviour (system with $N=14$ and PB). The most significant feature observed is the reduction in the total magnetisation with respect to the PB case due to the lower coordination at the surface. 
%----------------------------FIG.2b--------------------------------------------- 
\begin{figure}[tp] 
\centering 
\includegraphics[width= 1.0\textwidth]{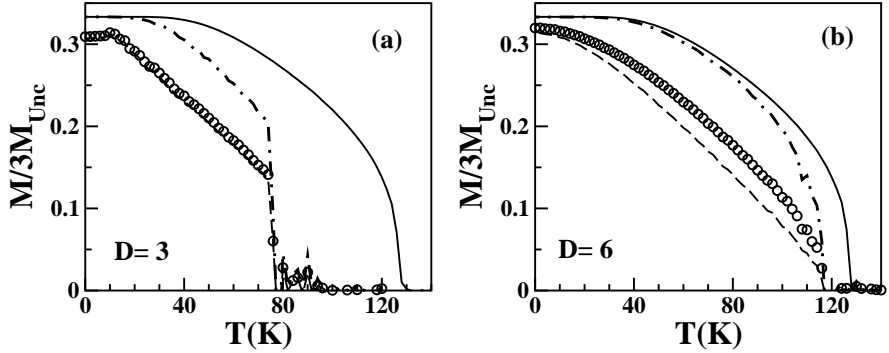}
\caption{
(a) Thermal dependence of the magnetisation $M$. The results for two particle diameters  
are shown: $D= 3, 6$ (left and right-hand panels respectively).  
The contributions of the surface (dashed line, for $D=3$ it cannot be distinguished from the total magnetisation) and core spins (dot-dashed line) have been distinguished from the total magnetisation (circles).  
The results for PB conditions, in a system of linear size  
$N= 14$, have also been included for comparison (continuous line).
}  
\label{M(T)_Ds_fig} 
\end{figure}
%----------------------------FIG.2b--------------------------------------------- 
Moreover, by plotting separately $M_{Core}$ and $M_{Surf}$ (see Fig. \ref{M(T)_Ds_fig}), we can differentiate the roles played by the surface and core in establishing the magnetic order. 
On the one hand, the core (dash-dotted lines) tends to attain perfect ferrimagnetic alignment ($M=1/3$) at low $T$ independently of the particle size, whereas for the surface spins (dashed lines) a rapid thermal demagnetisation is observed that makes $M_{Surf}$ depart from the bulk behaviour.
The behaviour of $M_{Total}$ is strongly dominated by the surface contribution for the smallest studied sizes ($D=3, 4$) and progressively tends to the bulk behaviour as the particle size is increased (see Fig. \ref{M(T)_Ds_fig}b). 
%----------------------------FIG.3--------------------------------------------- 
\begin{figure}[tp] 
\centering 
\includegraphics[width= 1.0\textwidth]{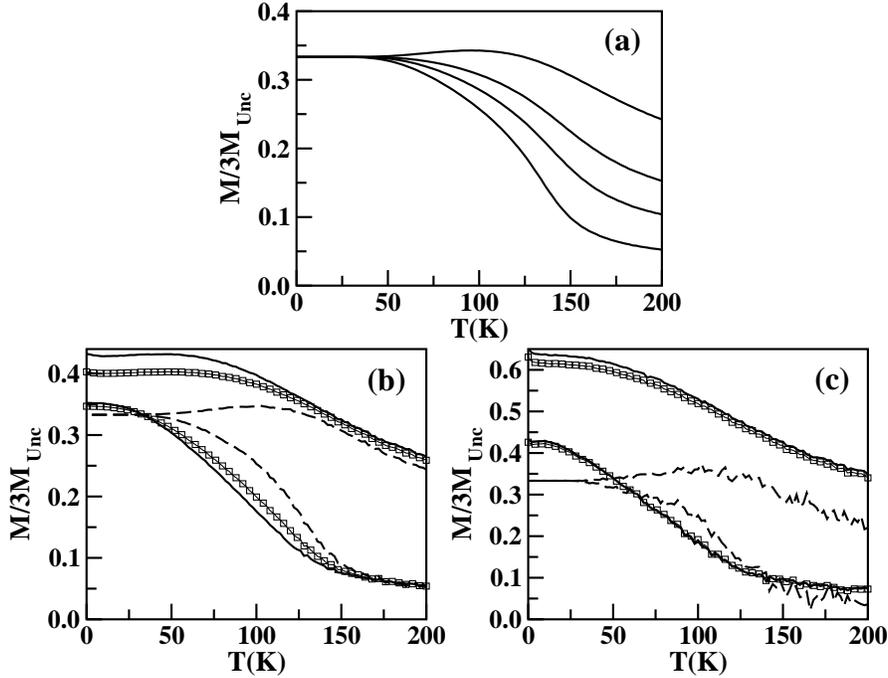}
\caption{
Thermal dependence of $M$ after cooling under $h_{FC}$. 
(a) Curves corresponding to a system with PB conditions and after cooling at $h_{FC}= 20, 40, 80, 100\ $K (from lower to uppermost curves).
(b) Corresponding curves for a spherical particle with $D= 3$. 
The results for two cooling fields $h_{FC}= 20\ $K (lower curve) and  
$h_{FC}= 100\ $K (upper curve) are shown. The contributions of the surface (thick lines) and the core (dashed lines) to the total magnetisation (squares) have been plotted separately.
(c) The same than in (b) for $D= 6$.  
}  
\label{MMM01_FC_fig} 
\end{figure}
%----------------------------FIG.3--------------------------------------------- 

Notice also that for all the studied sizes there is a region of linear dependence of the demagnetisation curve at intermediate $T$ that becomes wider as the particle size is decreased. In this region, the behaviour of the core is strongly correlated to that of the surface. 
This effect is indicative of a change from 3D to 2D effective behaviour of the surface layer of spin and has previously been observed in thin film systems \cite{Martinezjpcm92,Parkprl98} and in simulations of rough FM surfaces \cite{Zhaoprb00}. 

In contrast to the size dependence of the ordering temperature, the spontaneous magnetisation $M_{Total}$ follows, at any temperature, a quasi-linear behaviour with $1/D$ (see Fig. \ref{M(T)_Db_fig}b) indicating that the reduction of $M_{Total}$ is simply proportional to the ratio of surface-to-core spins, so it is mainly a surface effect. 
This is consistent with the existence of a surface layer of constant thickness $\Delta r$ independent of $D$ and with reduced magnetisation with respect to the core. With these 
assumptions, the size dependence of $M$ can be expressed as
\beq
M(D)=M_{Core}-\Delta M \frac{\Delta r S}{V}= M_{Core}- 
\Delta M \frac{6\Delta r}{D} \ ,
\eeq
where $S$ and $V$ are the surface and volume of the particle, and $\Delta M= M_{Core}-M_{Surface}$. 
Similar experimental behaviour has been found in $\gamma$-Fe$_2$O$_3$ \cite{Hanjm94}, microcrystalline BaFe$_{10.4}$Co$_{0.8}$Ti$_{0.8}$O$_{19}$ \cite{Batllejap93} and 
the above-mentioned MnFe$_2$O$_4$ system \cite{Tangprl91}.

%-------------------------------------------------------------------------------------
Deeper insight on the magnetic ordering of this system can be gained by studying 
the thermal dependence of the equilibrium magnetisation in a magnetic field. Several such curves obtained by the same cooling procedure as described previously in the presence of different fields $h_{FC}$ are presented in Fig. \ref{MMM01_FC_fig}. 
For particles of finite size, the curves at different fields do not converge to the ferrimagnetic value at low $T$, reaching higher values of the magnetisation at $T= 0$ the higher the cooling field (see lines with squares in Fig. \ref{MMM01_FC_fig}b,c). 
The total magnetisation for small particles is  
completely dominated by the surface contribution and this is the reason why the 
ferrimagnetic order is less perfect at these small sizes and the magnetic field can easily 
magnetise the system.
This is in contrast with the results for PB for which the system reaches perfect ferrimagnetic order (i.e. $M=1/3$) even in cooling fields as high as 100 K (see Fig. \ref{MMM01_FC_fig}a). Moreover, for this case, a maximum appears at high enough cooling fields $h_{FC}= 100\ $K,  which is due to the competition between the ferromagnetic alignment induced by the field and the spontaneous ferrimagnetic order (as the temperature is reduced, the strength of the field is not enough to reverse the spins into the field direction).

%However, the behaviour of the core is still  
%very similar to that of the case with PB, although its contribution  
%to $M$ is small.  
At low fields, the surface is in a more disordered state than the core since  
its magnetisation lies below $M$ at temperatures for which the thermal  
energy dominates the Zeeman energy of the field.
In contrast, a high field is able to magnetise the surface more easily than the  
core due to the fact that the broken links at the surface worsen the  
ferrimagnetic order while the core spins align towards the field  
direction in a more coherent way.   

%-------------------------------------------------------------------------------------
As a conclusion of the $M(T)$ dependence obtained in our simulations let us emphasize that, in spherical particles, there is a surface layer with much higher degree of magnetic disorder than the core, which is the Ising version of the random  canting of surface spins occurring in several fine particles with spinel structure \cite{Kodamaprl96,Linjm95,Coeyprl71,Morrishjap81,Jiangjpcm99}. 
However, our model does not evidence the freezing of this disordered layer in a spin-glass-like state below any temperature and size, in contrast with the suggestions given by some authors \cite{Martinezprl98,Kodamaprl97}
and some experimental findings \cite{Martinezprl98,Montseprb99,Gazeaueul97}, 
for instance, the observation of shifted loops after an FC process (see Fig. \ref{Fig5a_fig} and Refs. \cite{Montseprb99,,Moralescm99,Troncjm03}). 
%----------------------------FIG.4--------------------------------------------- 
\begin{figure}[tp] 
\centering 
\includegraphics[width= 1.0\textwidth]{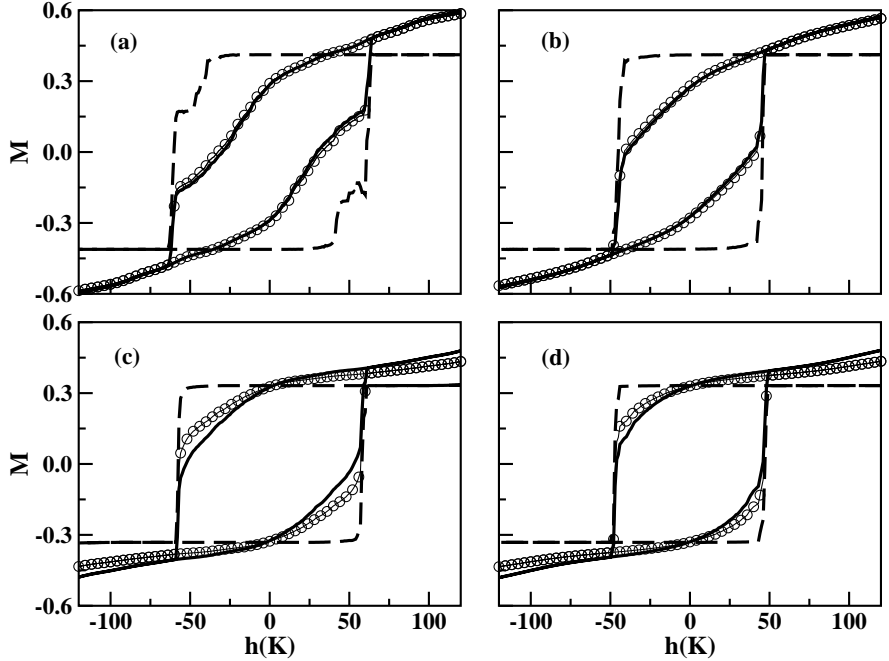}
\caption{Surface (continuous line), core (dashed line) and total (circles) contributions to the magnetisation of the hysteresis loops for particles of diameters $D= 3$, $T= 10$ K (a); 
$D= 3$, $T=20$ K (b); $D= 6$, $T= 10$ K (c); $D= 6$, $T=20$ K (d).
}  
\label{Fe2O3_HIDnv0SC_fig} 
\end{figure}
%----------------------------FIG.4--------------------------------------------- 
%
Furthermore, the surface layer, by partially breaking the ferrimagnetic 
correlations, diminishes the zero-field $M_{Total}$ but, 
at the same time, enhances $M_{Total}$ at moderate fields. 
Although the surface is easily thermally demagnetised and more easily magnetised 
by the field than the core is, it does not behave as a dead layer, since, 
at any $T$, it is magnetically coupled to the core. 
All these facts indicate that the surface has higher magnetic response 
than the core, excluding a spin-glass freezing. 
Moreover, we have not observed irreversibilities between field and 
zero-field cooled magnetisation curves in contrast to the experimental observations, 
which is a key signature that within the scope of our model, neither finite-size nor surface effects are enough to account for the spin-glass-like state. 
\subsection{Hysteresis loops}

More insight into the magnetisation processes can be gained by studying the thermal
and size dependence of hysteresis loops which, on the other hand, are the more common
kind of magnetic measurements. In order to simulate the loops, the system has been initially prepared in a demagnetised state at $h=0$ and the field subsequently increased in constant steps $\delta h= 1$ K up to a point well beyond the irreversibility field, from which the
field is cycled. Measurements of the different magnetisation contributions are performed at each step during $3000$ MCS and the results averaged for several independent runs.
%----------------------------FIG.4b--------------------------------------------- 
\begin{figure}[tp] 
\centering 
\includegraphics[width= 0.95\textwidth]{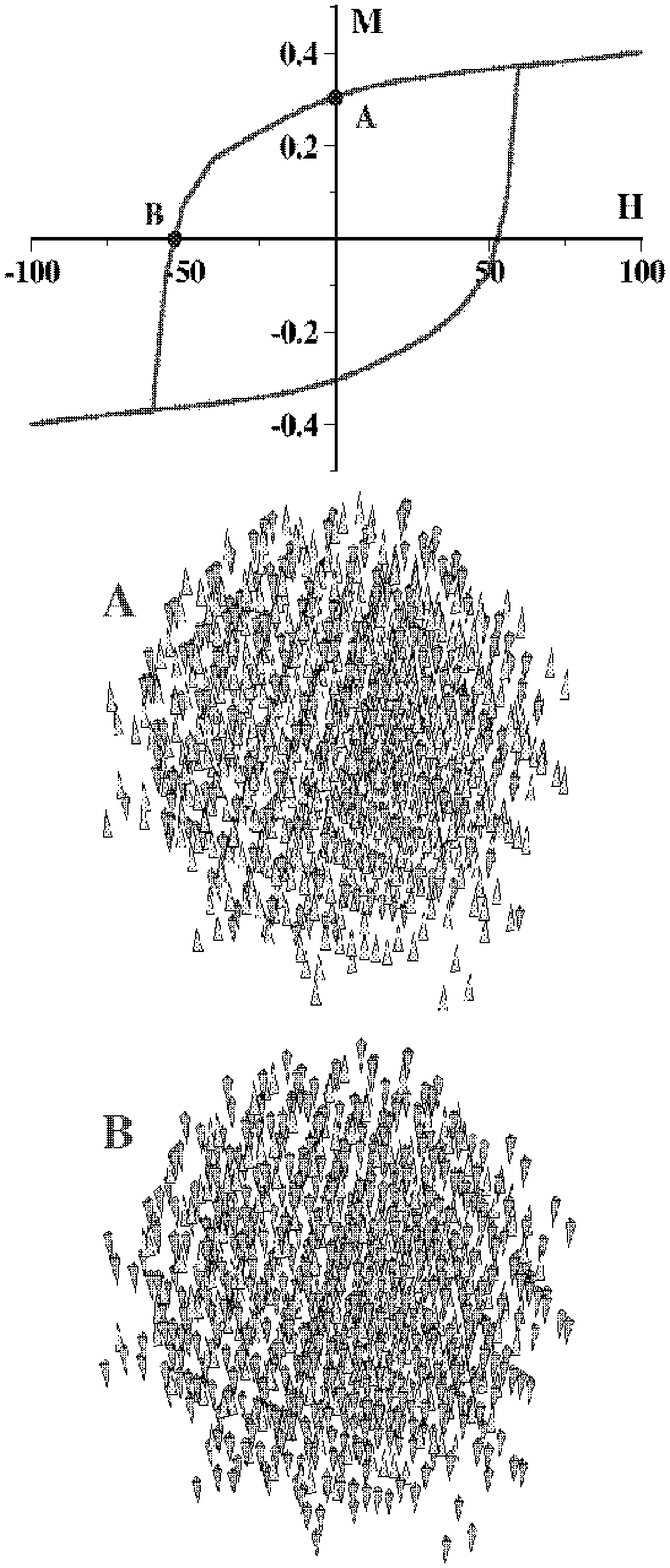}
\caption{Spin configurations of a particle of diameter $D=4$ at different points of the hysteresis loop for $T= 20$ K. Configuration A corresponds to the remanent magnetisation state whereas B corresponds to the zero magnetisation state. 
}  
\label{Reversal_fig} 
\end{figure}
%----------------------------FIG.4b--------------------------------------------- 

In Fig. \ref{Fe2O3_HIDnv0SC_fig}, we show the core and surface contributions to the hysteresis loops for two particles sizes $D=3,6$ and temperatures $T=10, 20$ K. 
First of all, let us notice that the shape of the loops for $M_{Total}$ reproduces 
qualitatively the main features observed experimentally: high field susceptibilities
and increasing saturation fields as the particle size is decreased. Both facts can be
attributed to the progressive alignment of surface spins towards the field direction. 
The loops of the smallest particles resemble those found in ferrimagnetic nanoparticles \cite{Kodamaprb99,Kodamaprl96,Kodamaprl97} and other bulk systems with disorder \cite{Binder,Maletta81}, increasing their squaredness 
(associated with the reversal of $M$ as a whole) with the size.

Secondly, the panels shown in Fig. \ref{Fe2O3_HIDnv0SC_fig} allow us to distinguish between the very different roles played by the surface and core in the reversal process. 
On the one hand, the particle core presents an almost perfect squared loop (dashed lines) independently of the particle size, which indicates that the core spins reverse in 
a coherent fashion so that, in spite of the AF inter- and intra-lattice interactions, one can consider the interior of the ferrimagnetic particle behaving as a single superspin with magnitude equal to the number of uncompensated spins. 
On the contrary, the surface contribution dominates the reversal behaviour of the particle 
(compare the continuous curves with those in circles) and
reveals a progressive reversal of $M$, which is a typical feature of disordered 
and frustrated systems \cite{Binder,Maletta81}.
Nonetheless, for a wide range of temperatures and particle
sizes, it is the reversal of the surface spins which triggers the reversal 
of the core. This is indicated by the fact that the coercive field of 
the core is slightly higher but very similar to the one of the surface. 
The surface spin disorder can be clearly observed in the snapshots of configurations taken along the hysteresis loop presented in Fig. \ref{Reversal_fig}, at the remanence point and at the coercive field (named A and B in the figure), where the different degree of disorder of the surface with respect to the core is more evident.
%%%%

The thermal dependence of the coercive field $h_c(T)$ for all the studied particle sizes, 
at difference with ferromagnetic particles, shows a complex behaviour mainly related to 
the frustrating character of the antiferromagnetic interactions and the non-trivial 
geometry of the spinel spin lattice. 
%----------------------------FIG.5--------------------------------------------- 
\begin{figure}[tp] 
\centering 
\includegraphics[width= 1.0\textwidth]{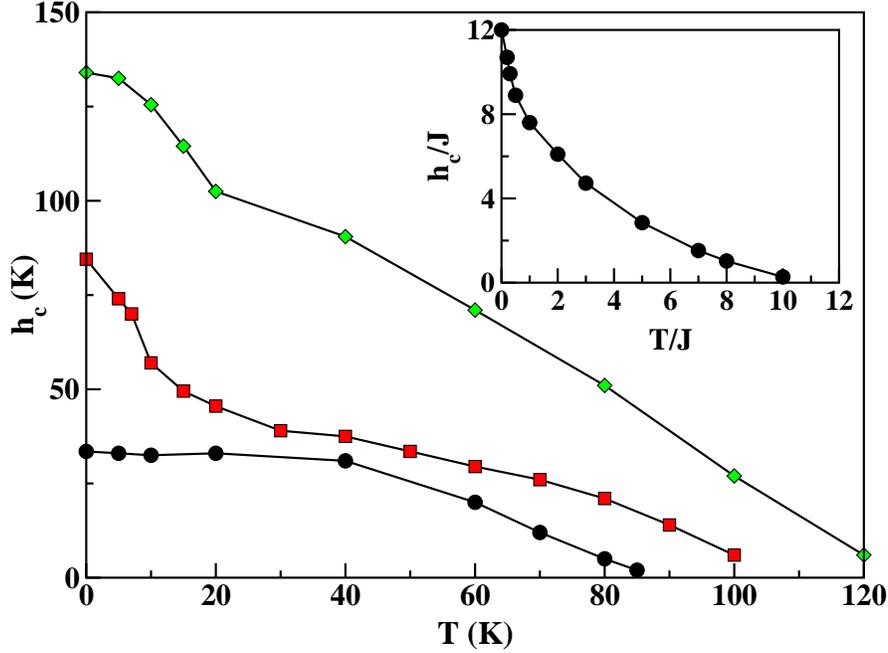}
\caption{
Temperature dependence of the coercive field $h_c$
for the real AF values of the exchange constants for maghemite for
the case of FB spherical particles of diameters 
$D= 3$ (circles), $D= 6$ (squares), and for a system of linear size
$N=8$ with PB conditions (diamonds).
Inset: Temperature dependence of the coercive field $h_c$ for a system with
the same structure as maghemite but ferromagnetic interactions 
($J_{\alpha\beta}=J$), PB conditions and $N= 8$.
}  
\label{Fe2O3_HCoeDnv0_fig} 
\end{figure}
%----------------------------FIG.5--------------------------------------------- 
%
Therefore, whereas simulations for a system with PB conditions and with the same structure as maghemite but equal interspin FM interactions show a monotonous decrease of $h_c$ with increasing $T$ (see the innset in Fig. \ref{Fe2O3_HCoeDnv0_fig}), the results for real maghemite display different thermal variations depending on the particle size (see Fig. \ref{Fe2O3_HCoeDnv0_fig}).
The FM case, $h_c(T)$ presents a power law decay for high enough temperatures
($T/J\gtrsim 1$), that can be fitted to the expression 
\begin{eqnarray}
	h_{c}(T)= h_{c}(0)[1-\left(T/T_{c}\right)^{1/\alpha}] \ , 
\label{Hcoe} 
\end{eqnarray}
with $\alpha= 2.26\pm 0.03$; close but different from what would be obtained 
by a model of uniform reversal such as Stoner-Wohlfarth \cite{Stoner48} 
($\alpha=2$). Thus, we see that, even in this simple case for which $M$ reverses as a whole, 
the thermal variation of $h_c(T)$ cannot only be ascribed to the
thermal activation of a constant magnetisation vector over an energy barrier 
landscape, since actually $M$ is of course temperature dependent.
%----------------------------FIG.6--------------------------------------------- 
\begin{figure}[tp]
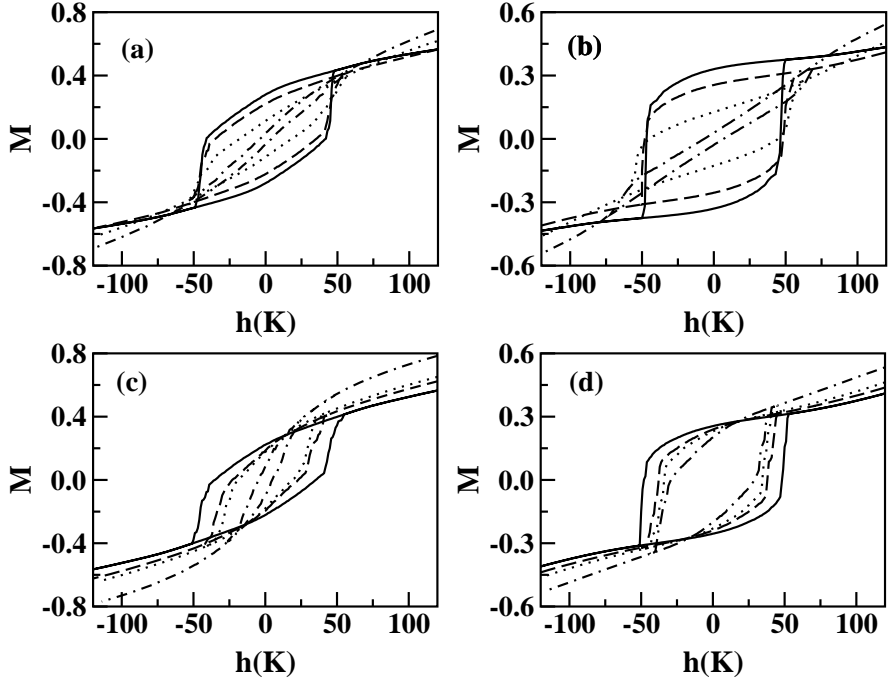
 
\centering 
\includegraphics[width= 1.0\textwidth]{Fig18a_HLoops_VacO.eps}
\includegraphics[width= 1.0\textwidth]{Fig18b_HLoops_VacSurf.eps}
\caption{
Upper panels:
hysteresis loops for systems with vacancy concentrations
$\rho_{v}$= 0.0, 0.166, 0.4, 0.6 (from outer to innermost) on the O sublattice at 
$T= 20$ K. Particle diameters $D=3$ (a), $D=6$ (b). Results have been averaged 
over 10 disorder realisations.
Lower panels:
hysteresis loops for systems with vacancy densities on the surface 
of the O and T sublattices $\rho_{sv}$= 0, 0.1, 0.2, 0.5, vacancy density 
$\rho_{sv}= 0.1666$ on the O sublattice, and $T= 20$ K.
Particle diameters $D=3$ (c), $D=6$ (d). 
Results have been averaged over 10 disorder realisations.
}  
\label{T10_Fig1114_fig} 
\end{figure}
%----------------------------FIG.6--------------------------------------------- 
%
In contrast with the FM case, the $h_c(T)$ dependence for maghemite particles with PB or spherical shape, presents opposite curvature and two regimes of variation. In the PB case at high $T$, data can also be fitted to the power law of Eq. \ref{Hcoe} for $T\gtrsim 20$ K with $\alpha = 0.94\pm 0.02$, $h_c(0)=134\pm 2$ K. Values of $\alpha$ close to 1 have been deduced in the past for some models of domain wall motion \cite{Gauntjap86,Gauntphilmagb83}. However, at low $T$, a different regime is entered but tending to the same $h_c(0)$ value. 
In this regime, the hysteresis loops become step-like and this change in behaviour is associated with the wandering of the system through metastable states with $M_{Total}\simeq 0$, which are induced by the frustration among AF interactions \cite{Iglesiasprb01}. 

Finally, the thermal dependence for spherical maghemite particles strongly depends on the particle size.
Notice that $h_c$ values are always smaller than the bulk (PB) values independently of $D$ because, for a finite particle, the existence of surface spins with reduced coordination acts as a seed for the reversal of the rest of the particle, which is not the case for PB, where all equivalent spins have the same coordination. 
This fact explains why the $h_c$ values for PB are only recovered at low $T$ in the limit of large particle size, although the decay of $h_c$ is similar to that of the bulk for high $T$.
In this regime, $h_c$ is dominated by the surface as indicated by the similarity of the core and surface $h_c$ already pointed out in Fig. \ref{Fe2O3_HIDnv0SC_fig}.
Instead, the low $T$ ($T\lesssim 20$) regime becomes dominated by the core contribution as 
$h_c^{Surf}<h_c^{Core}$ for any particle size (see Fig. \ref{Fe2O3_HIDnv0SC_fig}). However, for the $D=3$ particle, the prevalence of the core is hindered due to the small ratio of core-to-surface spins in this case, causing the saturation of $h_c$ when lowering the temperature.
\subsection{Effects of the disorder}

In real maghemite particles, disorder and imperfections cause the system to depart from perfect stoichiometry and distort the position of the atoms on the lattice, their effect being more important at the surface \cite{Moralesjpcm97}. 
There are several ways to implement this disorder on the model. 
The simplest way to simulate the deviation of the O and T sublattice atoms 
from ideal stoichiometry is by random removal of magnetic ions on the  
O and T sublattices. 

\subsubsection{Lattice disorder}

The ideal maghemite lattice presents one sixth of randomly distributed vacancies on the O sublattice in order to achieve charge neutrality that, however, in real samples, can vary 
depending on the conditions and method of preparation or the size of the particles \cite{Moralesjpcm97,Moralesjmr94,Sernassc01}. In the model presented so far, we have only considered
perfect lattices, now we will study the effect of vacancies on the magnetic properties.
Given the dominance of AF intersublattice interactions in maghemite, the inclusion of vacancies on one of the sublattices destabilises the perfect FM intralattice order of the other
and therefore may result in a system with a greater degree of magnetic disorder.
Since $N_{TO}>N_{OT}$ (see Section \ref{Frustration_Sec}), this effect will be stronger when vacencies are introduced in the O sublattice. In the following, we will refer to $\rho_{v}$ as the vacancy concentration on this sublattice.
%----------------------------FIG.7--------------------------------------------- 
\begin{figure}[tp] 
\centering 
\includegraphics[width= 1.0\textwidth]{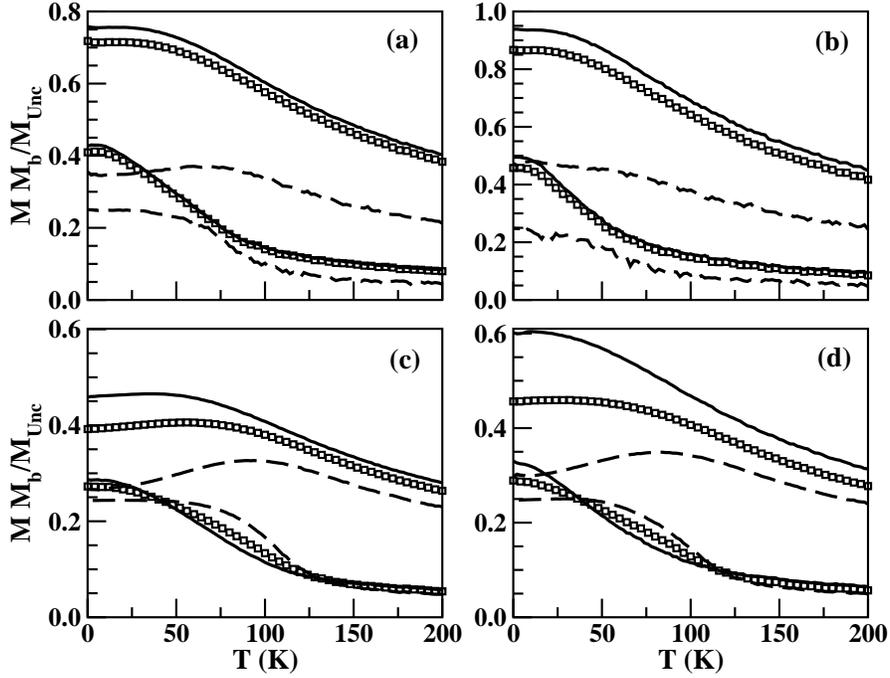}
\caption{Thermal dependence of $M$ after cooling under a magnetic field 
for a spherical particle with $D= 3$ (upper panels) and $D= 6$ (lower panels), 
with vacancy densities on the surface of the O and T sublattices $\rho_{sv}=$ 0.2 (a) and (c), $\rho_{sv}=$ 0.5 (b) and (d), and $\rho_{v}=$ 0.166 on the O sublattice.The results for 
two cooling fields $h_{FC}$= 20, 100 K (lower and upper curves respectively in 
each pannel) are shown. 
The contributions of the surface (thick lines) and the core (dashed lines) 
to the total magnetisation (squares) have been plotted separately.  
The magnetisation has been normalised to $M_{b}$, the magnetisation of
a perfect ferrimagnetic configuration for a system of infinite size.
}  
\label{FC_Disorder_fig} 
\end{figure}
%----------------------------FIG.7--------------------------------------------- 

To show the effect of these kinds
of disorder, we have simulated the hysteresis loops for different $\rho_v$ at two cooling fields $h_{FC}= 20, 100$ K. 
As can be seen in the upper panels of Fig. \ref{T10_Fig1114_fig},
the introduction of a low concentration of vacancies ($\rho_{v}=1/6$ as in
the real material) results in a reduction in the magnetisation and in an
increase of the high field susceptibility, without any substantial change 
in the general shape of the loops. 
However, if $\rho_{v}$ is increased 
beyond the actual value, the loops progressively close, loosing  
squaredness and progressively resembling those for a disordered system \cite{Binder,Maletta81}, 
with high values of the high field susceptibilities and much lower coercivity. 

\subsubsection{Surface disorder}

Let us now consider the effects of the disorder at the surface
of the particle, taking a $\rho_v = 1/6$ vacancy density on the O 
sublattice. Since the surface of the particles is not an ideal sphere, 
the outermost unit cells may have an increased number of vacancies on both  
sublattices with respect to those present in the core. Reduced coordination
at the surface may also change the number of links between the surface atoms. 
We will denote by $\rho_{sv}$ the concentration of surface vacancies
in the outermost primitive cells.

As for the case with no vacancies, the thermal dependence of the magnetisation 
in the presence of a magnetic field $h_{FC}$ helps to characterise the magnetic
ordering of the system. Several such curves are shown in Fig.  
\ref{FC_Disorder_fig}, in which the surface (continuous lines) and the core 
(dashed lines) contributions to the total magnetisation (open symbols) 
have been distinguished.
The introduction of vacancies does not change the low field behaviour of 
the total magnetisation, which is still dominated by the surface for both 
$D= 3, 6$, although the smallest particles are easily magnetised by the field.
However, at high fields, $M_{Total}$ is lower than $M_{Surf}$ and the surface 
progressively decouples from $M_{total}$ with the introduction 
of vacancies in the surface, this effect being more remarkable for the
biggest particle.
With respect to the core, at difference with the non-disordered case
($\rho_v = \rho_{sv}=0$), 
the low temperature plateau of $M_{Core}$ tends to a higher value than 
that for perfect ferrimagnetic order, since the main effect of the disorder 
is to break ferrimagnetic correlations in the core; increasing the 
ferromagnetic order induced by the field. This is reflected in a progressive
departure of the high and low field $M_{Core}$ curves with increasing 
disorder (see the dashed lines). 
The maximum appearing at high $h_{FC}$ is only slightly affected by disorder, 
shifting to lower $T$ and eventually disappearing for $D=3$ and $\rho_{sv}=0.5$.

Hysteresis loops with surface disorder are given in Fig. 
\ref{T10_Fig1114_fig}c,d for two particle diameters.
The introduction of surface vacancies facilitates the magnetisation 
reversal by progressive rotation, producing a 
rounding of the hysteresis loops when approaching $h_c$, in the same way 
as occurs when particle size is reduced.
The same fact explains the increase of the high field susceptibility, 
since the vacancies act as nucleation centers of FM domains 
at the surface, which, from there on, extend the FM correlations 
to the inner shells of spins. 
Moreover, a considerable decrease of $h_c$ is observed. All these facts
yield a progressive elongation of the loops, giving loop shapes 
resembling those of disordered systems \cite{Binder,Maletta81}. 
The lower panels in Fig. \ref{T10_Fig1114_fig}, where the surface and core contributions are shown separately, clearly evidence that the increase of FM 
correlations at the surface, facilitated by the vacancies, induces FM 
order in the core. 
That is to say, $M_{Core}$ follows the evolution of $M_{Surf}$ at moderate 
fields above $h_c$, in contrast with the case with no surface 
vacancies, where the core keeps the ferrimagnetic order for the same field range.   

\section{Open questions and perspectives}

We would like to conclude with an account of open questions and perspectives for future work. With respect to experiments, future work should be aimed at distinguishing which part of the effective energy barrier distribution in interacting fine-particle systems originates from the dipolar interactions and which one comes from frustration at the particle surface. In this respect, experiments in the spirit of those performed by Salling et al. \cite{Sallingjap94}, measuring the switching fields of individual ellipsoidal $\gamma$-Fe$_2$O$_3$ particles by Lorentz magnetometry, or by Wernsdorfer and co-workers \cite{Troncjm03,Jametprl01}, measuring magnetisation processes of a single particle by using microSQUIDs, should be encouraged. This would clarify the surface contribution. 
The synthesis of diluted samples with well-isolated particles, such as fluid suspensions or a solid matrix, would allow the study of shifted loop effects and the contribution of the surface to the irreversibility and closure fields.
In order to better characterise the intrinsic interaction effects on the formation of a collective state, the behavior of ferrimagnetic oxide systems could be compared to that of interacting metallic particles for which the surface glassy state is not observed. 

From the point of view of simulations it would be interesting to extend the results presented here for Ising spins to a model with Heisenberg spins with finite anisotropy.
This would allow one to account for the important role played by surface anisotropy on the magnetisation processes, a well-known phenomenon first described by N\'eel back in the 50's \cite{Neeljpr54} and that arises from the breaking of the crystalline symmetry at the boundaries of the particle.
This symmetry breaking changes also the effective exchange interactions at surface atoms with respect to the bulk values, a fact that could be easily incorporated in a model for MC simulation to study its possible influence on the appearance of shifted loops observed experimentally and that is not obtained within the scope of the Ising model.

\section{acknowledgments}
We are indebted to Dra. Montse Garc\'{\i}a del Muro for her contribution to some parts of the work reported in this chapter.
We also acknowledge CESCA and CEPBA under coordination of C$^4$ for the computer facilities. This work has been supported by SEEUID through project MAT2000-0858 and Catalan DURSI under project 2001SGR00066.

%%%%%%%%%%%%%%%%%%%%%%%%%%%%%%%%%%  BIBLIOGRAPHY %%%%%%%%%%%%%%%%%%%%%%%%%%%%%%%%%%%%%%%%%

%%%%%%%%%%%%%%%%%%%%%%%%%%%%%%%%%%  BIBLIOGRAPHY %%%%%%%%%%%%%%%%%%%%%%%%%%%%%%%%%%%%%%%%%
%\bibliographystyle{apalike}
%\bibliographystyle{prsty_noal}
%\bibliographystyle{natbib}
%{\bibliography{/Dades/Tesi/Refgen,/Dades/Tesi/Refmqt,/Dades/Tesi/Refmqtexp,/Dades/Tesi/Refpart,/Dades/Tesi/Refdip,/Dades/Tesi/Refmc,/Dades/Tesi/Refcamp,/Dades/Tesi/Refdipexp,/Dades/Tesi/Myreferences,/Dades/Tesi/Strings,/Dades/Tesi/Simulation}}
%\chapbblname{Review_bib_num}
%{\chapbibliography{/Dades/Tesi/Refgen,/Dades/Tesi/Refmqt,/Dades/Tesi/Refmqtexp,/Dades/Tesi/Refpart,/Dades/Tesi/Refdip,/Dades/Tesi/Refmc,/Dades/Tesi/Refcamp,/Dades/Tesi/Refdipexp,/Dades/Tesi/Myreferences,/Dades/Tesi/Strings,/Dades/Tesi/Simulation}}

\end{document}